\documentclass[aip, jcp, amsmath, amssymb,
 reprint,%
]{revtex4-1}

\usepackage{graphicx}
\usepackage{dcolumn}
\usepackage{bm}
\usepackage{booktabs}
\usepackage{color}

\newcommand{\vc}[1]{\mathbf{ #1}}    

\newcommand{\kT}{k_{\mathrm{B}}T}
\newcommand{\Nm}{N_{\text{m}}}

\newcommand{\rc}{r_{\text{c}}}
\newcommand{\Vm}{V_{\text{m}}}

\begin{document}


\title{Mesoscale simulations of confined Nafion thin films}

\author{P. Vanya}
 \email{peter.vanya@gmail.com}
 \affiliation{Department of Materials Science \& Metallurgy, University of Cambridge, 27 Charles Babbage Road,
 Cambridge CB3 0FS, United Kingdom}
 
\author{J. Sharman}
 \affiliation{Johnson Matthey Technology Centre, Blounts Court Road, Sonning Common, Reading RG4 9NH, United Kingdom}

\author{J.A. Elliott}
 \email{jae1001@cam.ac.uk}
 \affiliation{Department of Materials Science \& Metallurgy, University of Cambridge, 27 Charles Babbage Road,
 Cambridge CB3 0FS, United Kingdom}

\date{\today}

\begin{abstract}
The morphology and transport properties of thin films of the ionomer Nafion, with thicknesses on the order of the bulk cluster size, have been investigated as a model system to explain the anomalous behaviour of catalyst/electrode-polymer interfaces in membrane-electrode assemblies. We have employed dissipative particle dynamics (DPD) to investigate the interaction of water and fluorocarbon chains with carbon and quartz as confining materials for a wide range of operational water contents and film thicknesses. We found confinement-induced clustering of water perpendicular to the thin film. Hydrophobic carbon forms a water depletion zone near the film interface, whereas hydrophilic quartz results in a zone with excess water. There are, on average, oscillating water-rich and fluorocarbon-rich regions, in agreement with experimental results from neutron reflectometry. Water diffusivity shows increasing directional anisotropy of up to 30\% with decreasing film thickness, depending of the confining material. The percolation analysis revealed significant differences in water clustering and connectivity with the confining material. These findings indicate the fundamentally different nature of ionomer thin films, compared to membranes, and suggest explanations for increased ionic resistances observed in the catalyst layer.

\end{abstract}

\maketitle

%

\section{Introduction}
The discovery of Nafion, an ionomer used in hydrogen fuel cells, in the 1960s provided enormous potential for this clean technology to revolutionize the transport industry. Although, as of 2015, the first mass-produced fuel cell-powered cars have started to appear (e.g. the Toyota Mirai), several  problems, both theoretical and practical, are still restricting global-scale deployment. On the practical end there is the lack of hydrogen refuelling infrastructure, the relatively high cost, and low mechanical resilience of the membrane, if not suitably reinforced, in comparison with combustion engines. 

Nafion and other chemically similar ionomer membranes have received an enormous amount of scientific attention, as summarised in a number of review articles focusing on structure,~\citep{Mauritz_ChemRev_2004} transport properties,~\citep{Kreuer_ChemRev_2004, Weber_ChemRev_2004, Paddison_AnnuRev_2003} and overall behaviour~\cite{Kusoglu_ChemRev_2017}. In trying to understand the structure of the water-rich ionic domains in membranes, many different models have been proposed. The first was the cluster-network model of Hsu \& Gierke~\cite{Gierke_1981} Whilst unable to describe quantitatively the X-ray scattering from Nafion, this model captures the essential qualitative feature of nanophase separation between ionic and fluorocarbon phases with a characteristic length scale of 3 to 5~nm. Gebel \emph{et al.}~\cite{Rubatat_MacroM_2002} claimed that a fibrillar model, in which extended fluorocarbon chains decorated by ionic side groups and water, gave a better match to small-angle X-ray (SAXS) data. In 2008, Schmidt-Rohr proposed a parallel cylinder model,~\cite{SchmidtRohr_NMat_2008} claiming that the SAXS scattering curve best supports a system where the backbone forms cylinders a few Angstroms wide and a few hundred nanometres long. Subsequently, Kreuer refuted Schmidt-Rohr's model and argued for flat and narrow water domains.~\cite{Kreuer_AFM_2013} The discussion about the morphology is still far from resolved, but recent mesoscale modelling results tend to favour structures which more resemble ion-clustered models for bulk ionomer.~\cite{Elliott_SM_2011}

Most of the studies carried out so far have focused on the bulk Nafion. However, its role role within catalyst layers in real membrane-electrode assemblies is limited. The hydrogen oxydation and oxygen reduction reactions take place within the anode and cathode catalyst layers respectively, which are complex structures comprising catalyst nanoparticles, thin films of ionomer, and free space, through which reactant gases flow in and product water, unreacted gases, and water vapour flow out.

It has been found that Nafion confined into a thin film has a structure vastly different from the bulk. The effect of confinement on structure has been experimentally studied by the NIST group using X-ray and neutron reflectometry.  Dura \emph{et al.} observed lamellae in a hydrated Nafion thin film deposited on SiO$_2$ surface, but not on Pt or Au surfaces.~\cite{Dura_MA_2009} Similarly, DeCaluwe \emph{et al.} discovered oscillations in composition between layers rich in water and those rich in fluorocarbon groups, effectively producing a lamellar structure close to the Nafion-substrate interface.~\cite{DeCaluwe_SM_2014} Eastman showed that a significant change in properties occurs at film thickness below 60~nm.~\cite{Eastman_MA_2012} In a different study,~\cite{Kim_MA_2013} a Nafion thin film deposited on a silica substrate and explored using neutron reflectivity revealed an anisotropy in that water was ordered in layers parallel to the substrate. Modestino \emph{et al.} observed thickness-dependent proton conductivity in a Nafion thin film.~\cite{Modestino_MA_2013} All these studies suggest that the ionomer films within catalyst layers have very different structure and properties to bulk membranes. Interpretations based on the assumption that catalyst layer ionomer behaves like very thin membrane material are likely to be misleading.

On the modelling side, many simulations have been done on bulk membranes using various theoretical frameworks.\cite{Elliott_PCCP_2007} A commonly used technique is dissipative particle dynamics (DPD), a mesoscale method capable of capturing significant length and time scales at a fraction of the computational cost of classical molecular dynamics. Its key features are a simple interparticle potential controlled by only one parameter and fast equilibration. This enables significant coverage of the parameter space at a reasonable computational cost, while retaining a system size of several tens of nanometres, a length scale necessary to observe the effects of water clustering. The most prominent studies have been carried out by Yamamoto \& Hyodo~\cite{Yamamoto_PolymJ_2003}, Wu \emph{et al.}~\cite{Wu_EES_2008, Wu_MA_2009, Wu_LA_2010}, and Dorenbos et al.~\cite{Dorenbos_JCP_2011, Dorenbos_JCP_2013, Dorenbos_JCP_2015}. All the workers were been able to reproduce structures qualitatively resembling the cluster-network model, with water cluster size of several nanometres, demonstrating that mesoscale methods offer a reliable insight into complicated polymer-solvent systems.

In comparison to bulk ionomers, there have been far fewer attempts to model a thin film version. Kendrick \emph{et al.} compared IR spectroscopy of a Nafion thin film deposited on a Pt \{111\} surface with DFT calculations.~\cite{Kendrick_JACS_2010} Nouri-Khorasani used classical molecular dynamics to calculate hydronium ion distribution and self-diffusion of water in a nanochannel~\cite{NouriKhorasani_2014} as well as in a film on a PtO substrate.~\cite{NouriKhorasani_2016} Borges used the same method to gain insight into the hydrophobicity of Nafion surface.~\cite{Borges_JPCC_2015,Borges_ACSN_2013}

On the mesoscale level, Dorenbos \emph{et al.} simulated Nafion confined by carbon surfaces at various hydrophobicities, which were controlled by the solubility parameter,~\cite{Dorenbos_ElComm_2010} revealing anisotropy in water diffusion, with a greater tendency for water to move parallel to the film. The authors claimed that the increased hydrophobicity of the carbon increased this anisotropy. 

In this paper, we explore a Nafion thin film via simulation, aiming to mimic the structure formed within the catalyst layer at the correct length scales. We investigate the changes in water morphology and transport properties for a wide range of operational water contents, from a very dry state to the membrane effectively immersed in water, and for a range of film thicknesses. For the confining substrates, we chose carbon and quartz as two opposite extremes in terms of hydrophobicity. Carbon is well-established as a fuel cell electrode material; quartz is often used as a substrate in neutron reflectometry studies.~\cite{DeCaluwe_SM_2014, Dura_MA_2009, Kim_MA_2013}

The remainder of the paper is structured as follows. First, we briefly review the DPD method and parameterisation procedure for ionomer thin film simulations, before presenting results on water distribution, diffusivity and connectivity. We then relate these results to the effects of confinement as a thin film on ionomer structure.

\section{Simulation Method}
\subsection{Overview of the method}
Dissipative particle dynamics (DPD) is a coarse-grained method using soft potentials bead-spring model of polymers to accelerate the dynamics. Introduced by Hoogerbrugge and Koelman~\cite{Hoogerbrugge_EPL_1992, Koelman_EPL_1993} to simulate suspension flows, DPD was reformulated by Groot and Warren (GW)~\cite{Groot_JCP_1997} to make it applicable to soft matter systems. GW set the method on a firm theoretical footing and created a protocol for calculating interaction parameters based on macroscopic compressibility and the Flory-Huggins theory.

In the past 20 years, DPD has been extensively applied to multiple families of soft matter systems, most notably diblock copolymers, surfactant solutions, and bilayer and ionomer membranes.~\cite{Groot_JCP_1998, Wu_EES_2008, Wu_MA_2009, Wu_LA_2010, Yamamoto_PolymJ_2003} For a summary of some applications we recommend consulting references~\cite{Karttunen_book_2003,Espanol_JCP_2017}.

The key idea behind fast dynamics of DPD is to combine a few molecules into a \emph{bead}, which is then used as a simulation particle. All beads should have approximately the same mass $m$ and radius $\rc$, thus enabling the use of reduced units $m=\rc =1$. The potential between the beads is quadratic:
\begin{equation}
V(r) = \begin{cases}
a (1-r)^2 \,/\, 2, & r < 1,\\
0,                 & r > 1,
\end{cases}
\end{equation}
where $a$ is the interaction parameter. The density is constant, usually between 3 and 5 beads per unit volume $\rc^3$. The method is defined in the NVT ensemble and the temperature is controlled by the Langevin thermostat.

Besides conceptual simplicity, DPD preserves hydrodynamics, i.e. the motion of beads resembles the fluid dynamics given by the Navier-Stokes equations.
This is an important advantage over e.g. the Brownian dynamics as another representative of coarse-grained methods.~\cite{Groot_JCP_1999} Among the well-documented disadvantages of DPD is inaccurate prediction of dynamic quantities.~\cite{Karttunen_book_2003}

GW showed that this simple form of interaction potential leads to a quadratic equation of state for $\rho\geq 3$: $p = \rho k T + \alpha a \rho^2$, where the proportionality constant $\alpha = 0.101$. The simplicity of the equation of state means that DPD is not able to reproduce complex phenomena such as vapour-liquid coexistence. The interaction parameter $a$ is the only free parameter to account for the differences between the beads. It can be split into the default value $a_0$, which can be derived from water compressibility, and excess repulsion $\Delta a_{ij}$ linearly proportional to the Flory-Huggins $\chi$-parameter, where $i,j$ are particle types.

\subsection{Parametrisation}
In DPD, only three forces exist: conservative due to the interparticle potential, and dissipative and random due to the Langevin thermostat:
\begin{equation}
\frac{d\vc v_i}{dt} = \vc f_i,\quad 
\vc f_i = \sum_{i\neq j} (\vc F^{\rm C} + \vc F^{\rm D} + \vc F^{\rm R}),
\end{equation}
where
\begin{align}
\vc F^{\rm C}(\vc r_{ij}) = &
\begin{cases}
a(1 - r_{ij}) \hat{\vc r}_{ij}, & r_{ij} < 1, \\
0,                              & r_{ij} > 1,
\end{cases}\\
\vc F^{\rm D}(\vc r_{ij}) = & \; - \gamma w^2(r_{ij})
(\hat{\vc r}_{ij} \cdot \vc v_{ij}) \hat{\vc r}_{ij}, \\
\vc F^{\rm R}(\vc r_{ij}) = & \;\sigma w(r_{ij}) \theta_{ij} \hat{\vc r}_{ij},
\end{align}
with $\vc v_{ij} = \vc v_i - \vc v_j$, $\vc r_{ij} = \vc r_i - \vc r_j$, $\hat{\vc r} = \vc r/|\vc r|$ and $\theta_{ij}$ being a Gaussian random number with zero mean and unit variance.

Espa\~{n}ol and Warren~\cite{Espanol_EPL_1995} showed that one has the freedom to choose the weight function $w(r)$ in the dissipative and random term. A simple linear choice is often made:
\begin{equation}
w(r) = \begin{cases}
(1-r), & r < 1,\\
0,     & r > 1.
\end{cases}
\end{equation}
These authors also related $\sigma$ and $\gamma$ to enforce the Boltzmann distribution onto the system: $\sigma^2 = 2\gamma \kT$. Following GW as well as other works, we choose $\sigma$ to be 3.0.

The default interaction parameter  $a$ for a bead containing one molecule is $25\,\kT$, which can be derived from water compressibility.\cite{Groot_JCP_1997} F\"uchslin \emph{et al.} have shown that this value remains scale-invariant in reduced units, i.e. if water serves as the solvent, $a=25\kT$ should always be used regardless of the coarse-graining degree (number of molecules in a bead). The coarse-graining degree only affects the way the quantities of interest are converted to the SI units after the simulation.

The cross interaction terms $\Delta a_{ij}$ between unlike beads are derived from the Flory-Huggins $\chi$-parameter. The DPD equation of state can be matched to the Flory-Huggins theory.\cite{Groot_JCP_1997} The relationship between $\Delta a$ and $\chi$-parameter is linear: $\Delta a = 3.27 \chi$ at density $\rho = 3$. The scaling of $\Delta a$ with the degree of coarse-graining remains debatable, and to date it is not clear whether $\Delta a$ should scale at all (see the discussion section in F\"uchslin \emph{et al.}~\cite{Fuchslin_JCP_2009}). For this reason, we chose to leave the excess repulsions unscaled.

Several methods are available to calculate the $\chi$-parameters. Common choices are a simple relation based on molar volume $\Vm$ and Hildebrand solubility parameters $\delta_i$ of component $i$:
\begin{equation}
\chi_{ij} = \frac {V_\text{m}} {RT} (\delta_i - \delta_j)^2,
\label{eq:sol}
\end{equation}
or more sophisticated Monte Carlo sampling developed by Fan \emph{et al.}~\cite{Fan_MA_1992}. We chose the mixture of these approaches: taking the $\chi$~parameters derived by Wu \emph{et al.} via the Monte Carlo sampling where available, and employing the more approximate eq~\eqref{eq:sol} in case of the interactions containing the substrate beads.

\subsection{System under investigation}
We set the DPD length scale $\rc$ as follows: starting from the approximate volume of one water molecule $V_0 = 30$~\AA$^3$, number of water molecules per bead $\Nm = 6$ and DPD density $\rho = 3$, the bead diameter for our simulation is $\rc = (V_0 \Nm \rho)^{1/3} = 8.14$~\AA. The box size is $40\times 40\times 40$ (in DPD units), corresponding to 32.5~nm in SI units and accommodating in total 192,000 beads.

The time scale $\tau = \sqrt{m \rc^2/\kT}$ is 5.35~ps. The simulation step was set to $\Delta t = 0.04\,\tau$. One simulation ran for 40,000 steps (8.6~ns), from which the first 30,000 served as equilibration. A few nanoseconds of equilibration are sufficient for a DPD simulation due to the softness of the interparticle potential.

The coarse-graining degree is six water molecules per bead. Common DPD parametrizations involve three~\cite{Groot_BiophysJ_2001} or four~\cite{Yamamoto_PolymJ_2003} water molecules per bead. Following Wu \emph{et al.}~\cite{Wu_EES_2008} we coarse-grained further in order to simulate larger boxes at the same computational overhead. In contrast with the workers,~\cite{Dorenbos_JMS_2009, Yamamoto_PolymJ_2003} we also put three water molecules in the C bead containing mainly the backbone and the sulfonic acid group to represent the strong binding between the acidic group and water. Some water is thus bound to polymer chains and its movement is more restricted. The water content $\lambda$ is defined as the number of water molecules per sulfonic acid group $N_{\rm{H_2 O}}/N_{\rm{SO_3 H^+}}$ with respect to the whole simulation box.

\begin{figure}
    \centering
    \includegraphics[width=8cm]{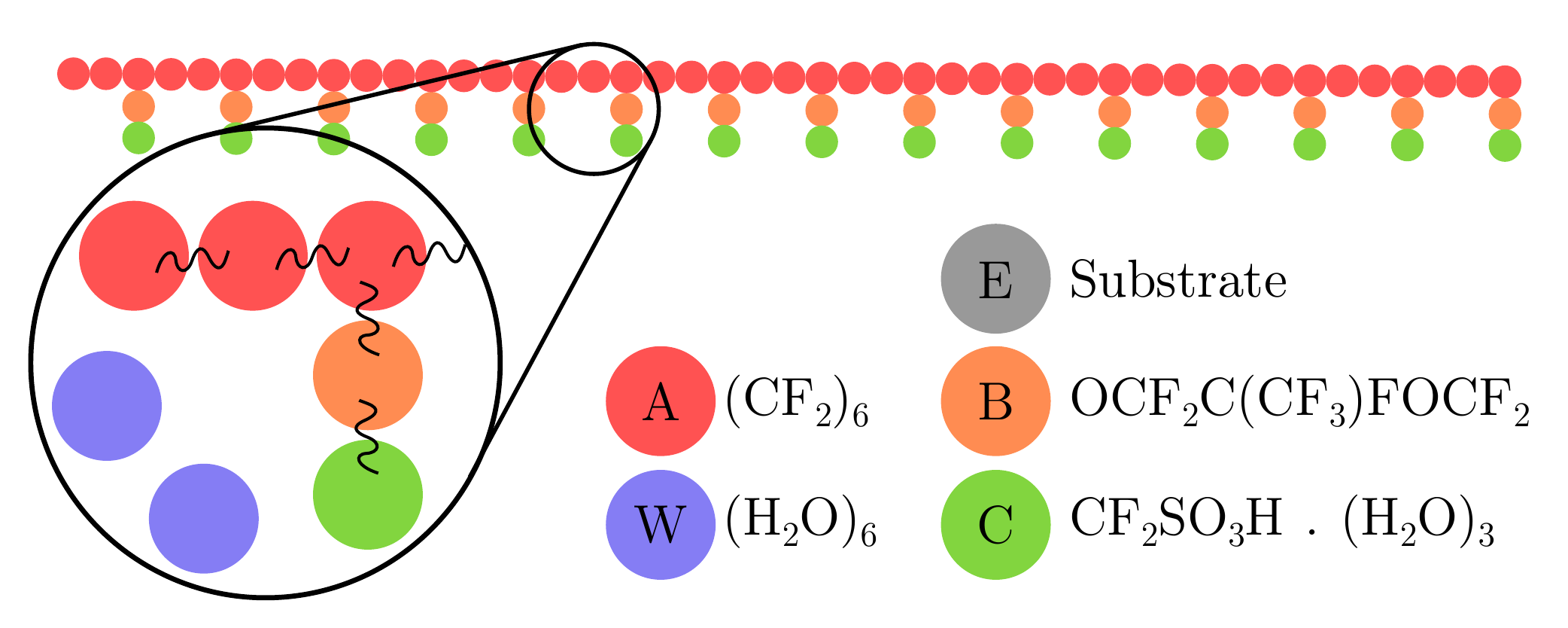}
    \caption{DPD beads used in the simulation. Each polymer chain has 15 segments, each segment has five beads. Water beads (blue) are freely floating around.}
    \label{fig:dpd_beads}
\end{figure}

The polymerisation of the PTFE chain is 15 (Fig.~\ref{fig:dpd_beads}) and each segment has five beads. Overall our polymer-solvent system contains five bead types: A, B, C, W, and E. To get the $\chi$-parameters For A, B, C, and W beads, we used the data from Wu \emph{et al.},~\cite{Wu_EES_2008}, who employed the method by Fan \emph{et al.}~\cite{Fan_MA_1992} 
Otherwise, we used eq.~\eqref{eq:sol}. The solubilities of A, B, and C beads were taken from Dorenbos~\cite{Dorenbos_ElComm_2010} and those for water, graphite and quartz were derived from the cohesive energy density (Table~\ref{tbl:params}). The molar volume of water in molecular form was considered for all beads $\Vm=18$~cm$^3\,$mol$^{-1}$, since all DPD beads are expected to have similar size and mass. The full cross-interaction parameter is:
\begin{equation}
a_{ij} = 25\,\kT + 3.27 \chi_{ij}.
\end{equation}

To understand and quantify the behaviour of confined Nafion we measured multiple characteristics: distribution of water in the direction parallel to the film, clustering of water in the film using percolation theory, and self-diffusion coefficient of the W beads.

The simulations were run in the DL\_MESO package.~\cite{DLMS} Box initialisation as well as post-processing of water distribution, diffusivity, and clustering were done using home-made Python scripts. Two confining materials with four film thicknesses and the bulk, plus eleven water contents, amounts to 99 different configurations in total. Each configuration was averaged over 3 to 6 independent trials differing by the random seed used for the box initialisation. The confining substrate beads were frozen for convenience during the simulation, since the substrate is solid and thus not expected to move on the same time scale as the thin film.

\begin{table}
\small
\setlength{\tabcolsep}{5pt}
\begin{ruledtabular}
\begin{tabular}{p{1.8cm}|c|llll}
           & $\delta$ (MPa$^{1/2}$) &\multicolumn{4}{c}{$\chi$ (no units)} \\
           & & A & B & C & W \\\hline
A          & 12.7 & 0  \\
B          & 13.6 & 1.23 & 0 \\
C          & 23.0 & 7.44 & 2.70 & 0 \\
W          & 47.8 & 3.36 & 1.53 & 1.48 & 0    \\
E (Carbon) & 25.0 & 1.10 & 0.94 & 0.03 & 3.77 \\
E (Quartz) & 35.0 & 3.60 & 3.32 & 1.04 & 1.19 \\
\end{tabular}
\end{ruledtabular}
\caption{Flory-Huggins $\chi$-parameters defined between pairs of beads used in the simulation. Excess repulsions $\Delta a_{ij} = 3.27\chi_{ij}$ were added to the default value $a=25\,\kT$.}
\label{tbl:params}
\end{table}

\subsection{Dynamics}
\label{sec:dynamics}
We calculated the diffusivity via the mean square displacement (MSD) defined as
\begin{equation}
D_d = \frac {\langle |\vc r(t_{\rm f}) - \vc r(t_{\rm i})|^2 \rangle } {2\, d\, (t_{\rm f} - t_{\rm i})},
\end{equation}
where $d\in\{1,2,3\}$ is the dimensionality of the system. It is known that polymer systems possess different regimes of diffusion at various time scales. We chose the initial and final times $t_{\rm i}$ and $t_{\rm f}$ respectively to so as to capture the linear regime, which we identified by plotting the MSD on a logarithmic scale. For carbon confinement, this regime takes place between 0.2 and 2$\tau$ and for quartz between 2 and 19$\tau$.

As a check we also recalculated the diffusivity via the velocity autocorrelation function, a different method using bead velocities instead positions: $D_\text{3d} = 1/3 \int_0^\infty \langle \vc v(0) \cdot \vc v(t) \rangle dt$. These two approaches yield the same results up to fluctuations, so one can remain confident in exploiting the MSD route.

DPD is known to overstate the dynamical properties due to its very soft force field. Tto put the water diffusion in a polymeric system into a meaningful perspective we compared all our calculations to the diffusivity $D_0=0.3$ of a pure liquid at standard simulation parameters mimicking pure water: $a=25\,\kT$, $\rho=3$, and $\gamma=4.5$.

\begin{figure*}
\centering
\includegraphics[width=0.38\columnwidth]{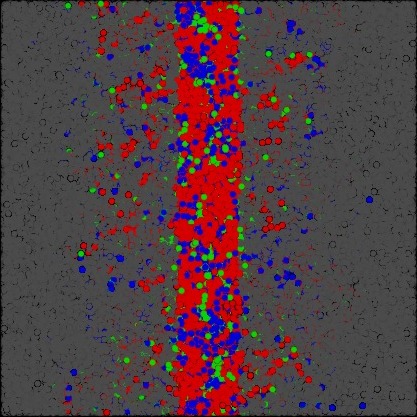}
~
\includegraphics[width=0.38\columnwidth]{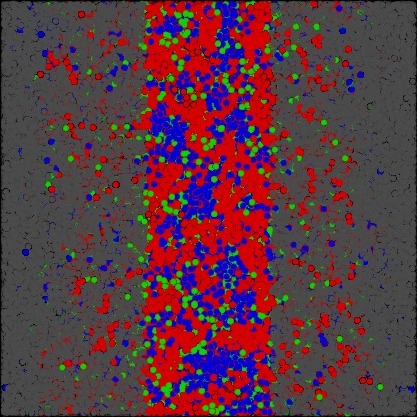}
~
\includegraphics[width=0.38\columnwidth]{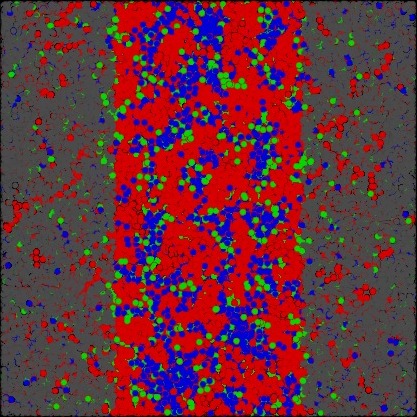}
~
\includegraphics[width=0.38\columnwidth]{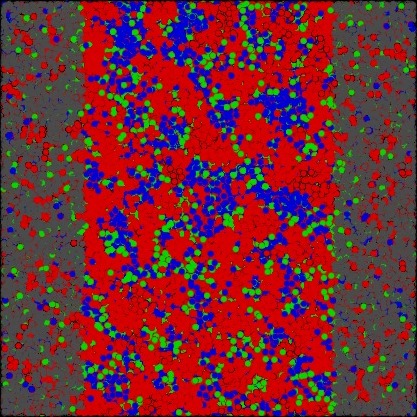}
~
\includegraphics[width=0.38\columnwidth]{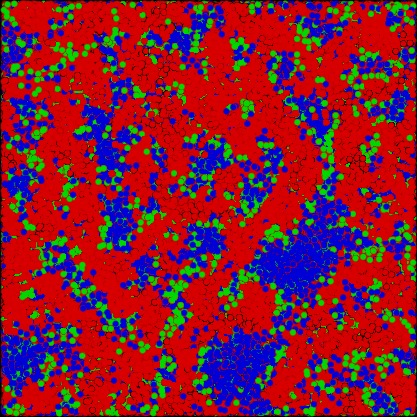}
\caption{(Color online) Equilibrated boxes of Nafion thin film confined by carbon for thicknesses $d = 5,10,15,20$~nm, respectively and the bulk, at water content $\lambda=16$. Red: backbone (A and B beads). Green: sulfonic acid groups (C). Blue: water (W). Grey: electrode (E).}
\label{fig:vmd}
\end{figure*}

\section{Results}
\begin{figure*}
\centering
\includegraphics[width=0.48\columnwidth]{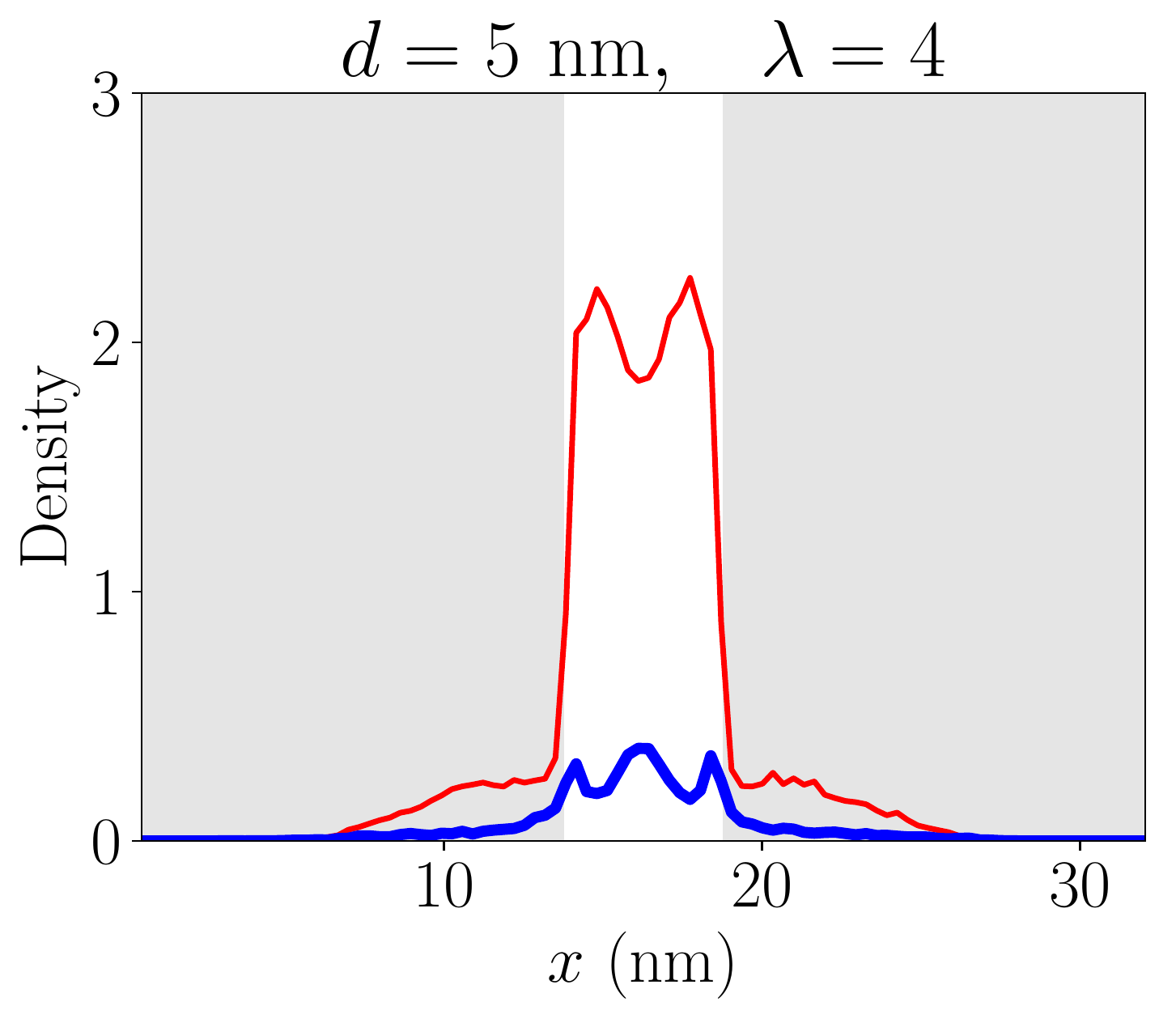}
~
\includegraphics[width=0.48\columnwidth]{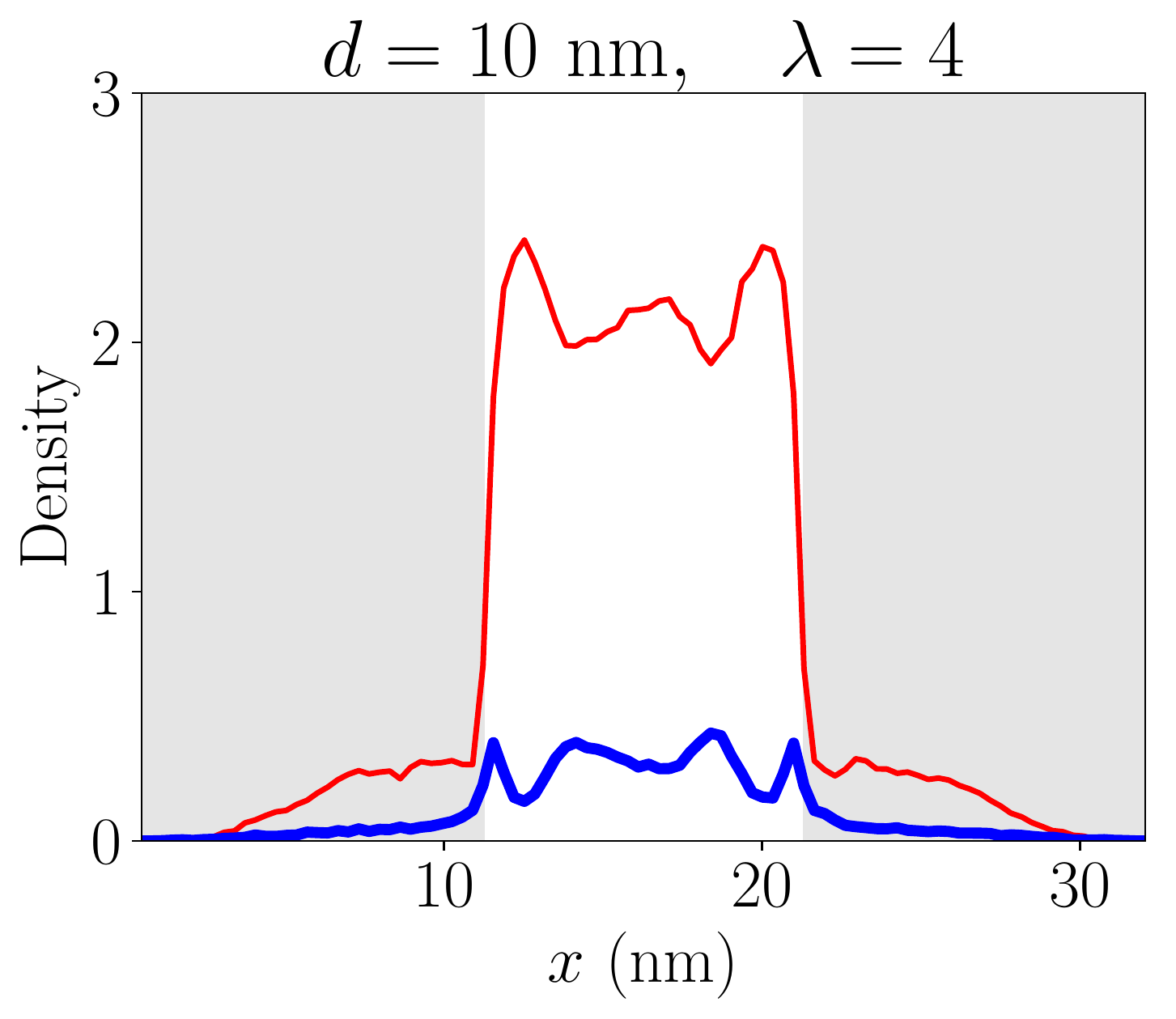}
~
\includegraphics[width=0.48\columnwidth]{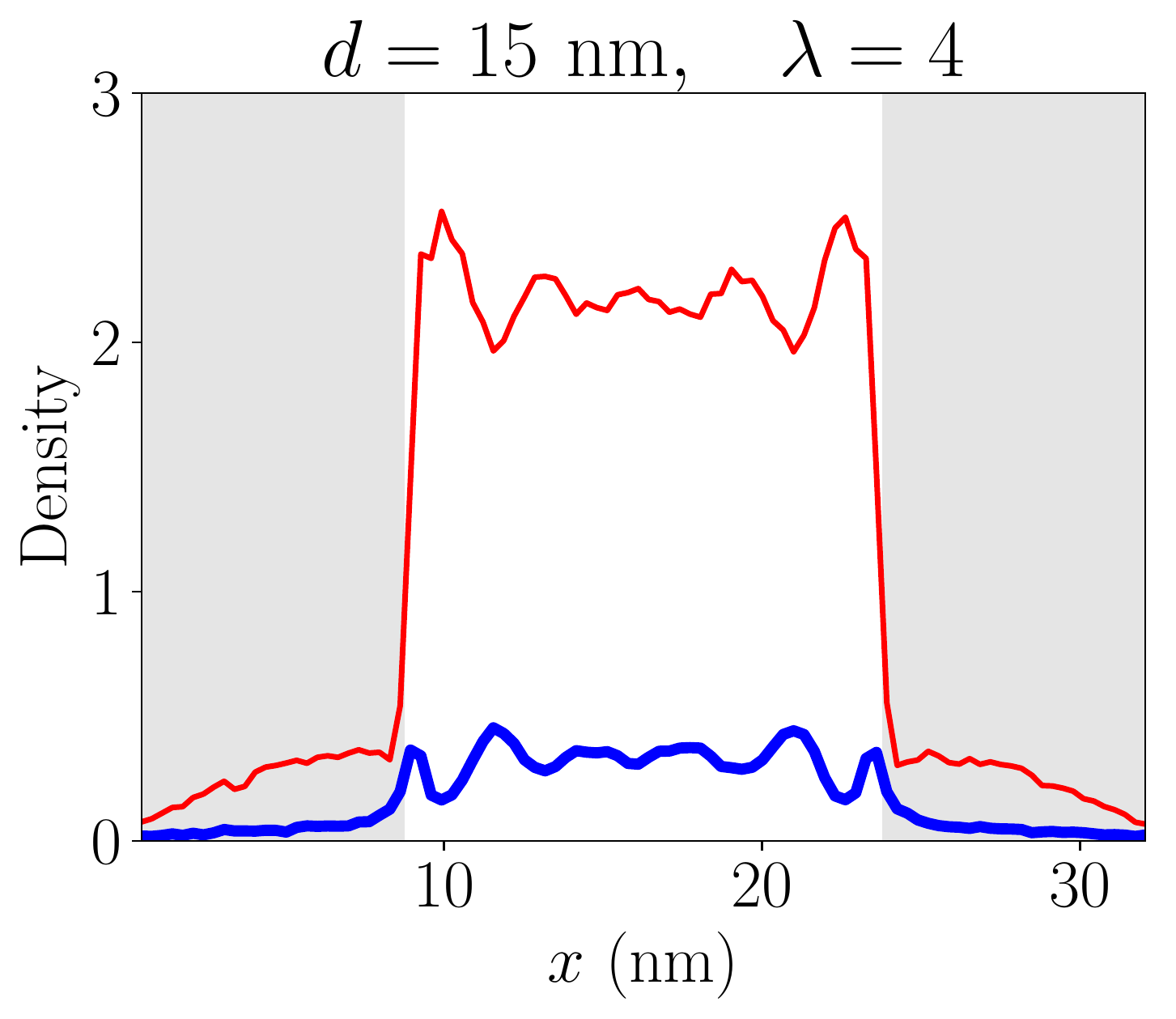}
~
\includegraphics[width=0.48\columnwidth]{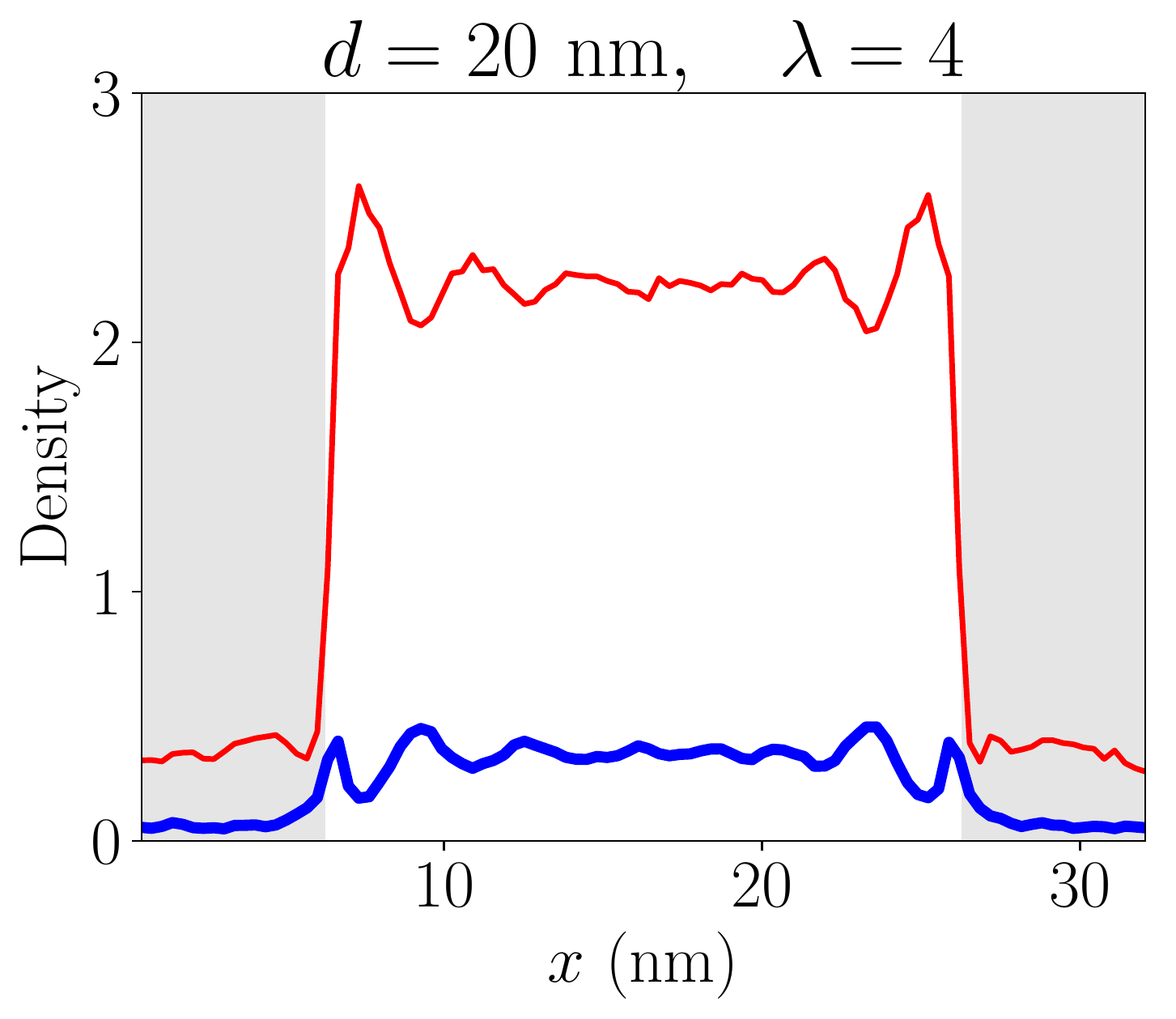}
\\
\includegraphics[width=0.48\columnwidth]{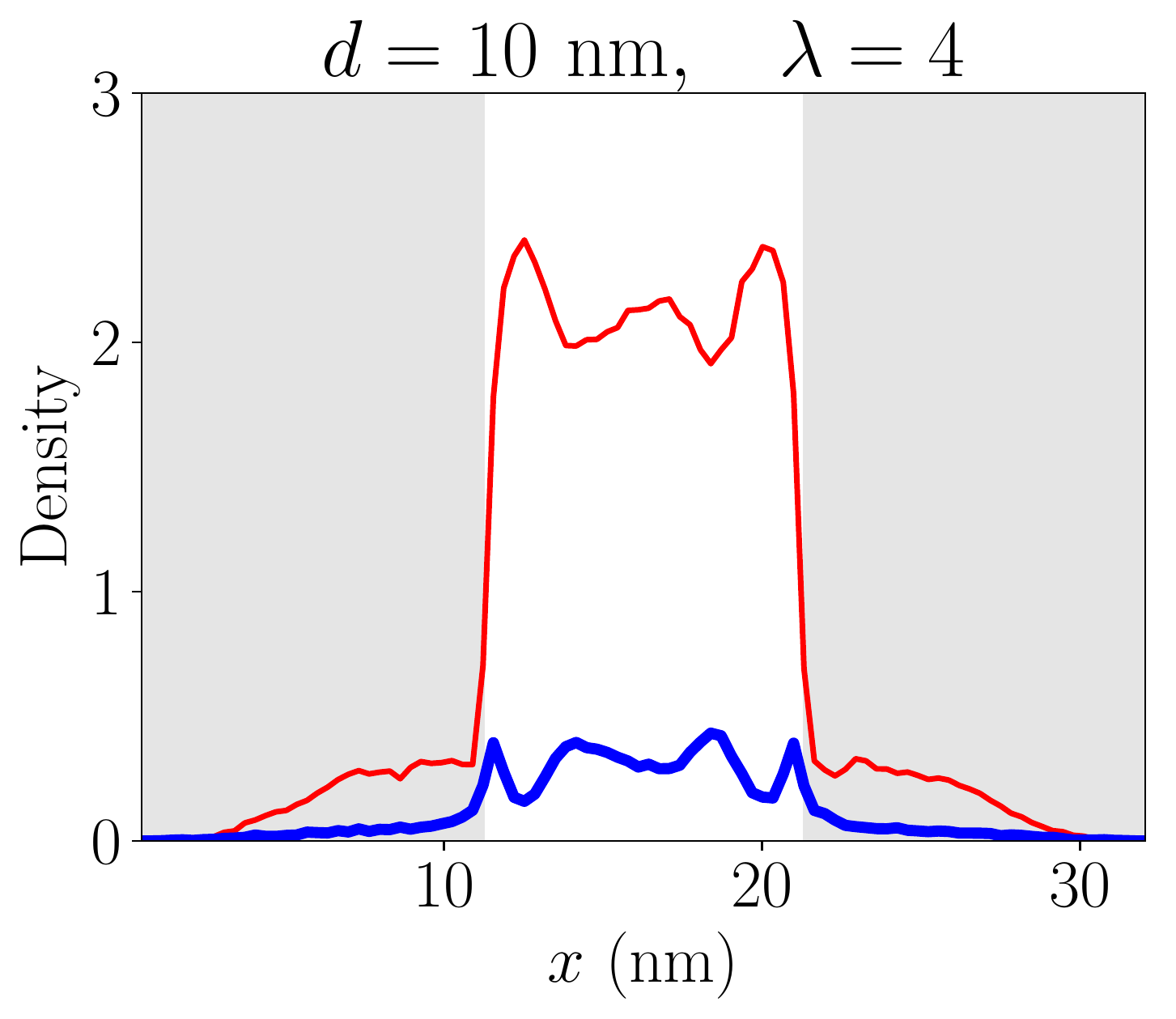}
~
\includegraphics[width=0.48\columnwidth]{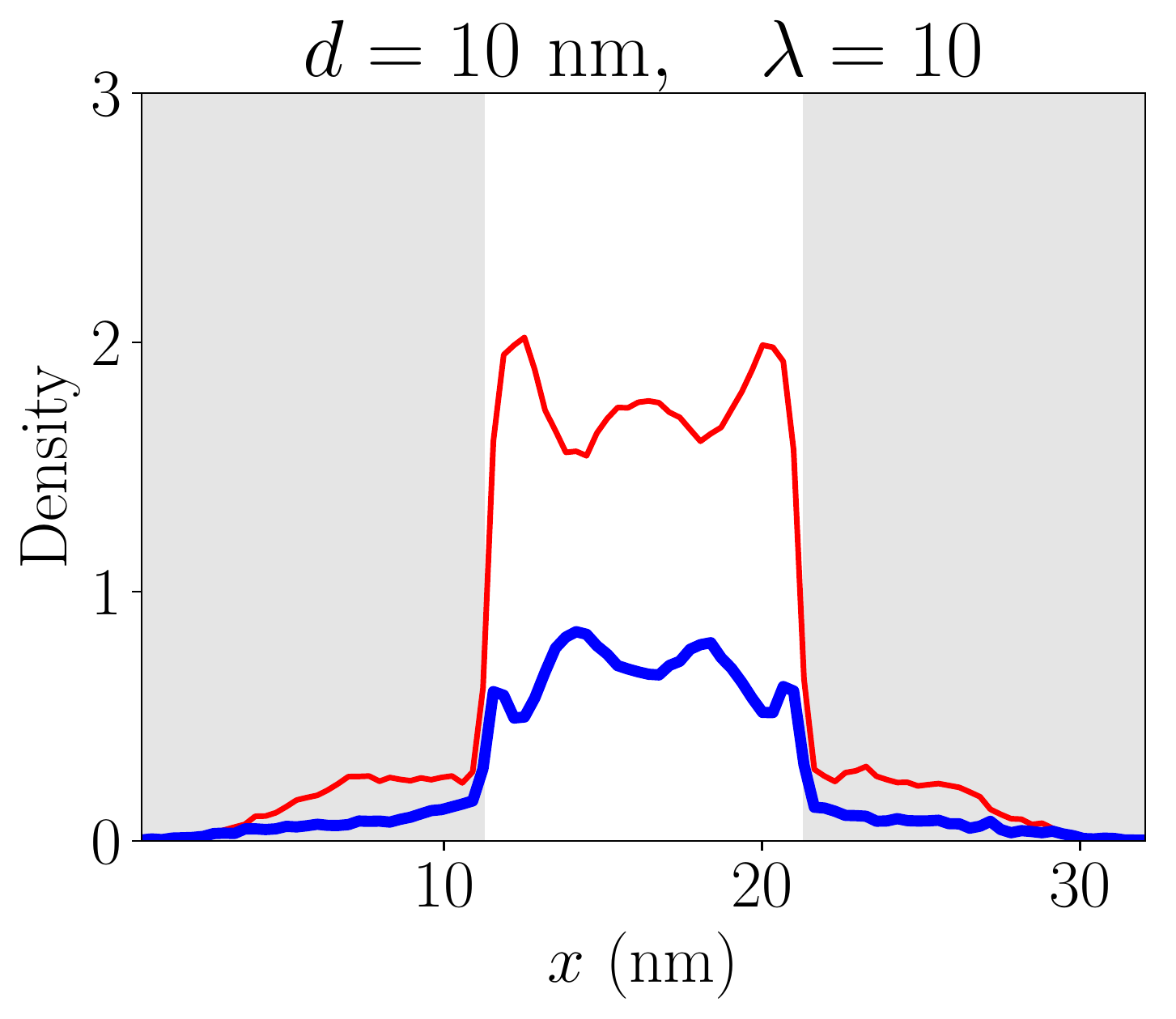}
~
\includegraphics[width=0.48\columnwidth]{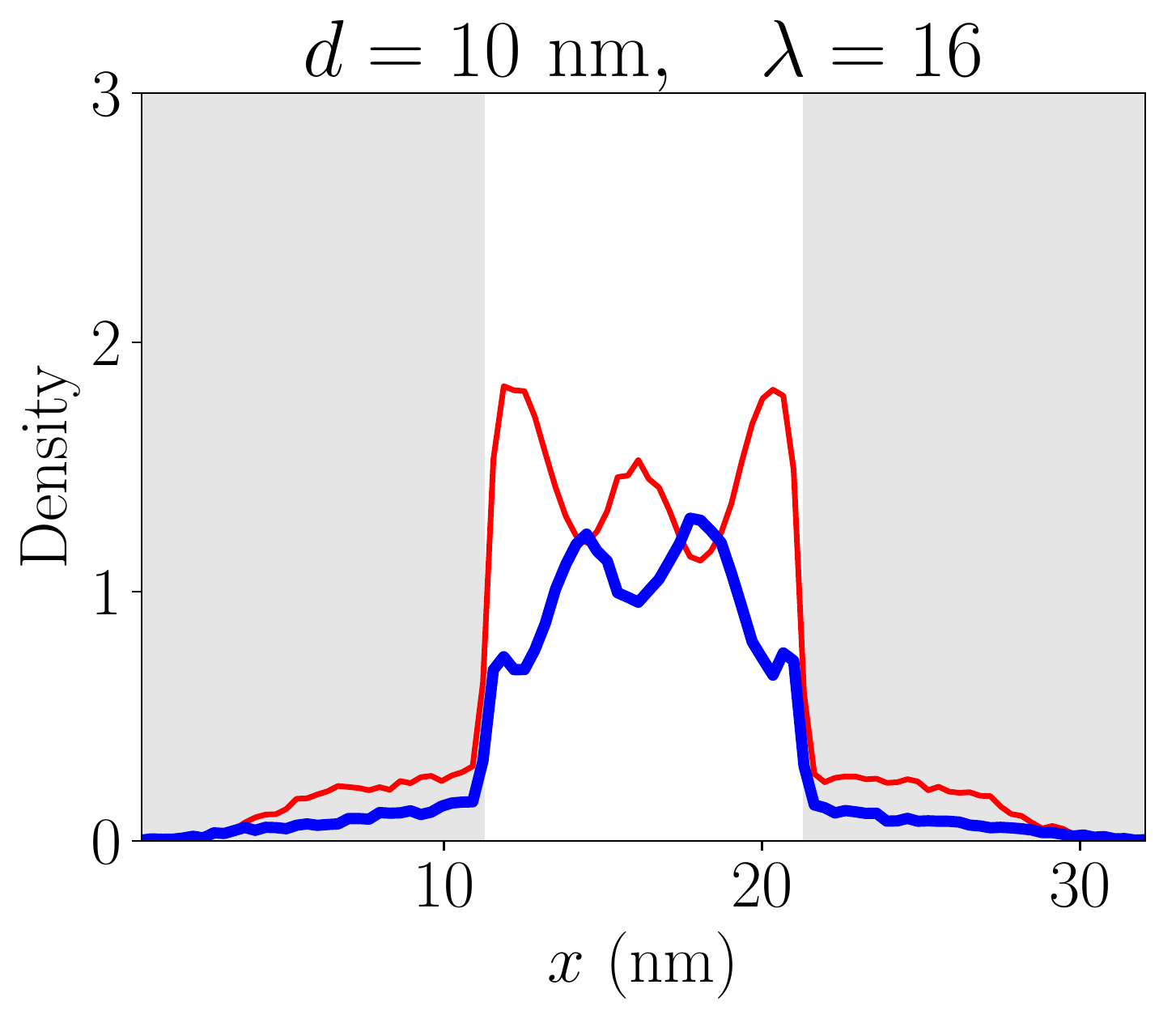}
~
\includegraphics[width=0.48\columnwidth]{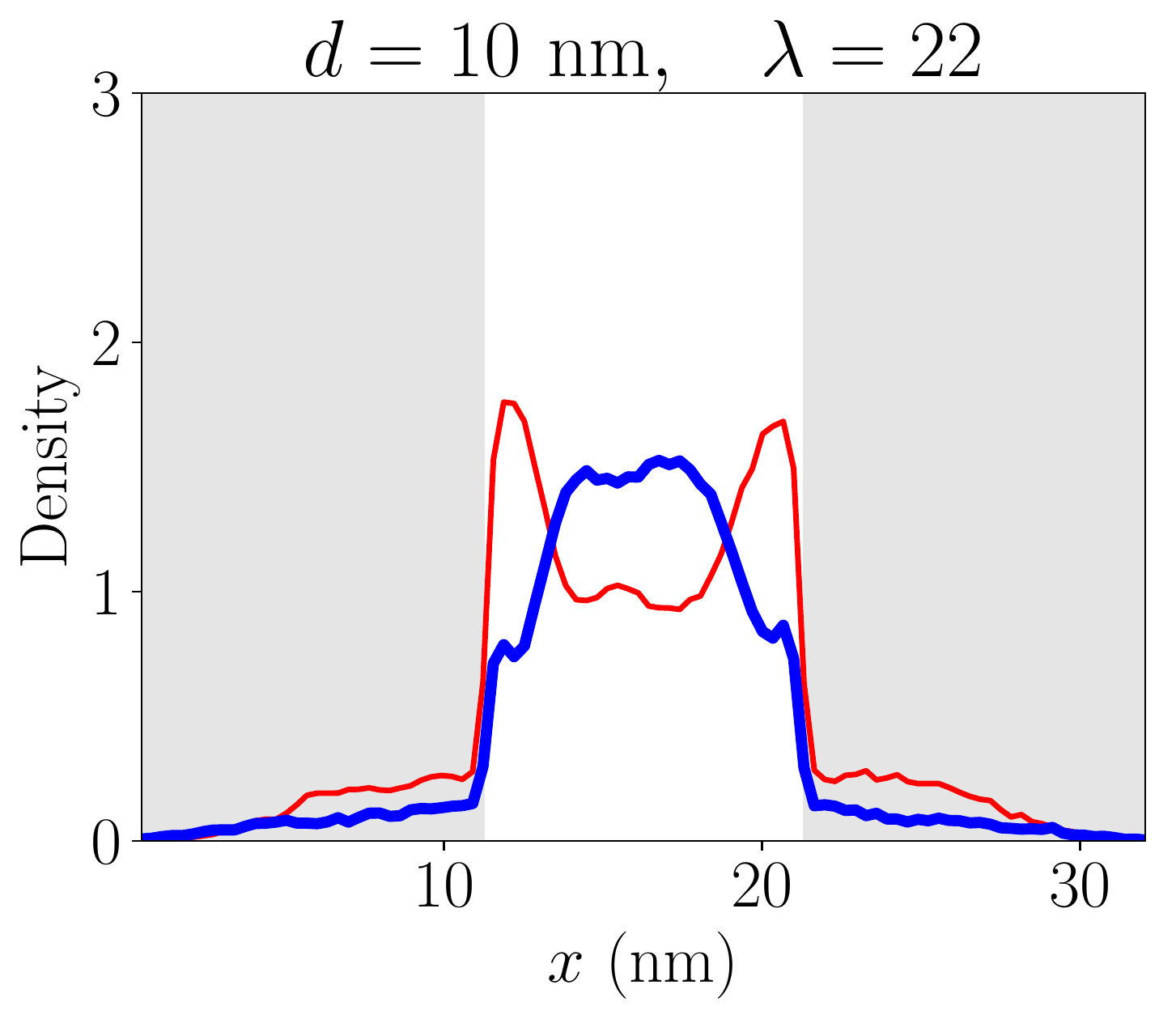}
\caption{(Color online) Top row: Profiles of water (blue) and PTFE backbone (red) confined by carbon for a range of film thicknesses $d = 5,10,15,20$~nm at water content $\lambda=4$.
Bottom row: Profiles of water (blue) and PTFE backbone (red) confined by carbon for a range of water contents $\lambda = 4,10,16,22$ at film thickness $d=10$~nm.}
\label{fig:profiles_carbon}
\end{figure*}

\begin{figure*}
\centering
\includegraphics[width=0.48\columnwidth]{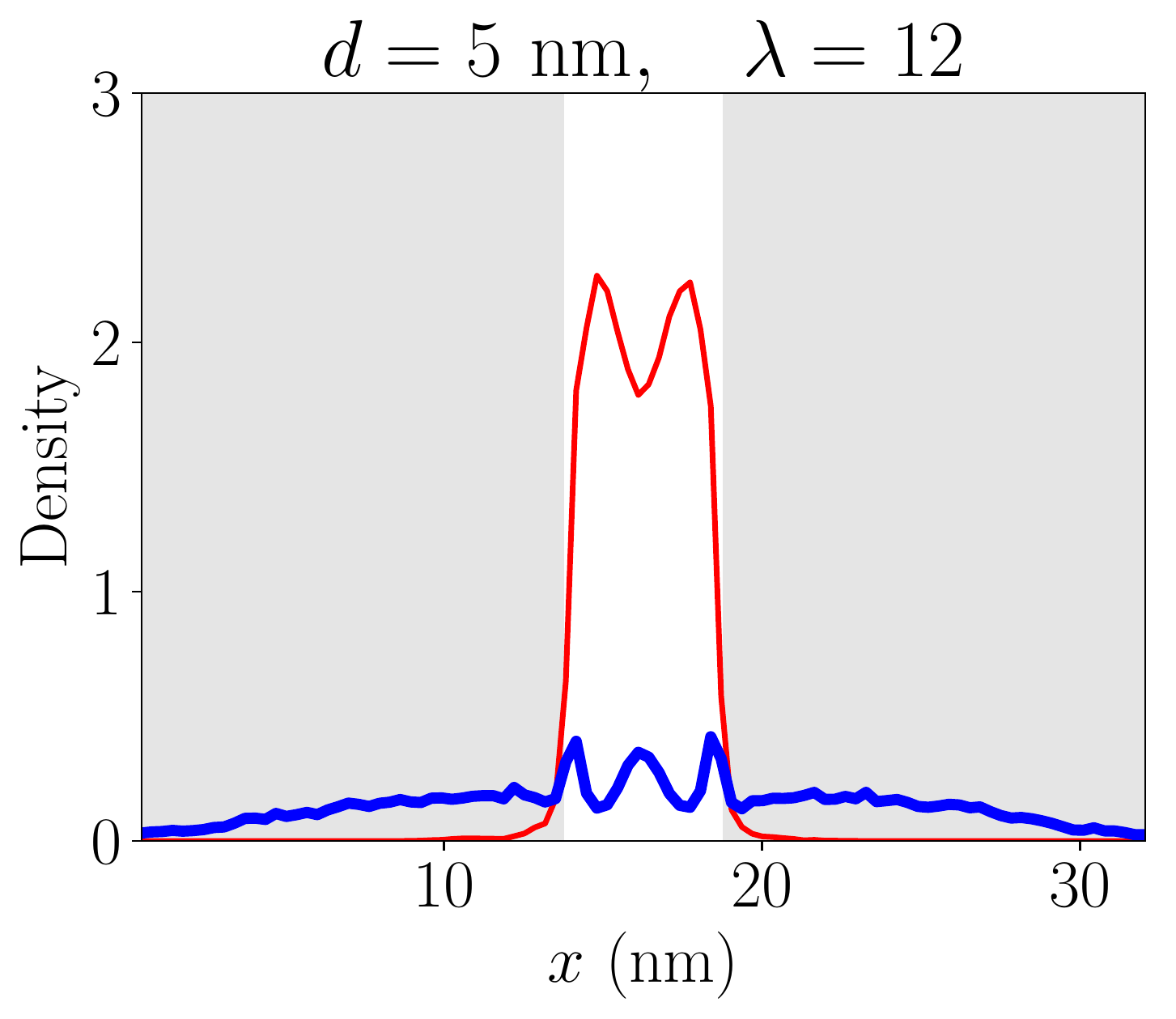}
~
\includegraphics[width=0.48\columnwidth]{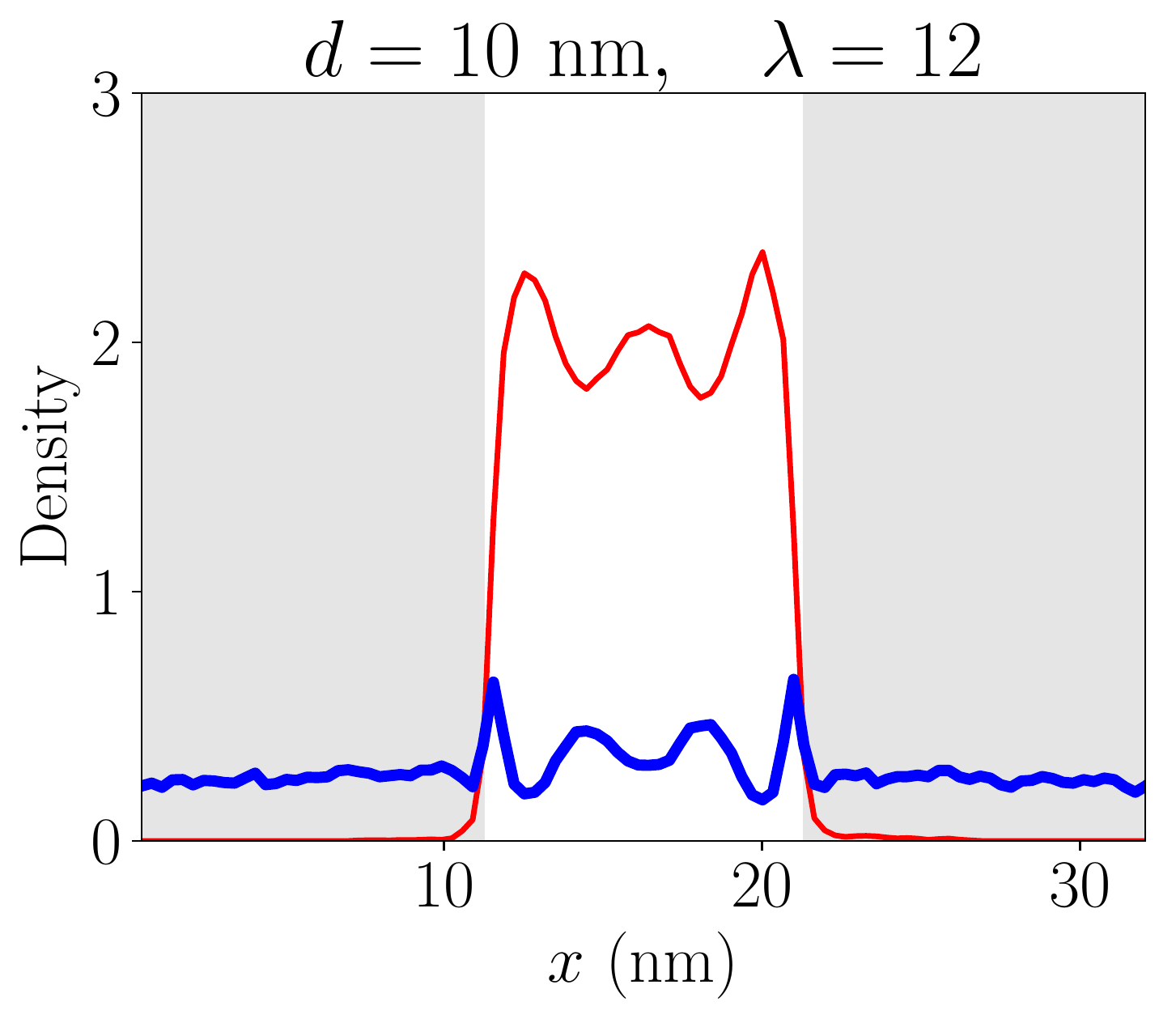}
~
\includegraphics[width=0.48\columnwidth]{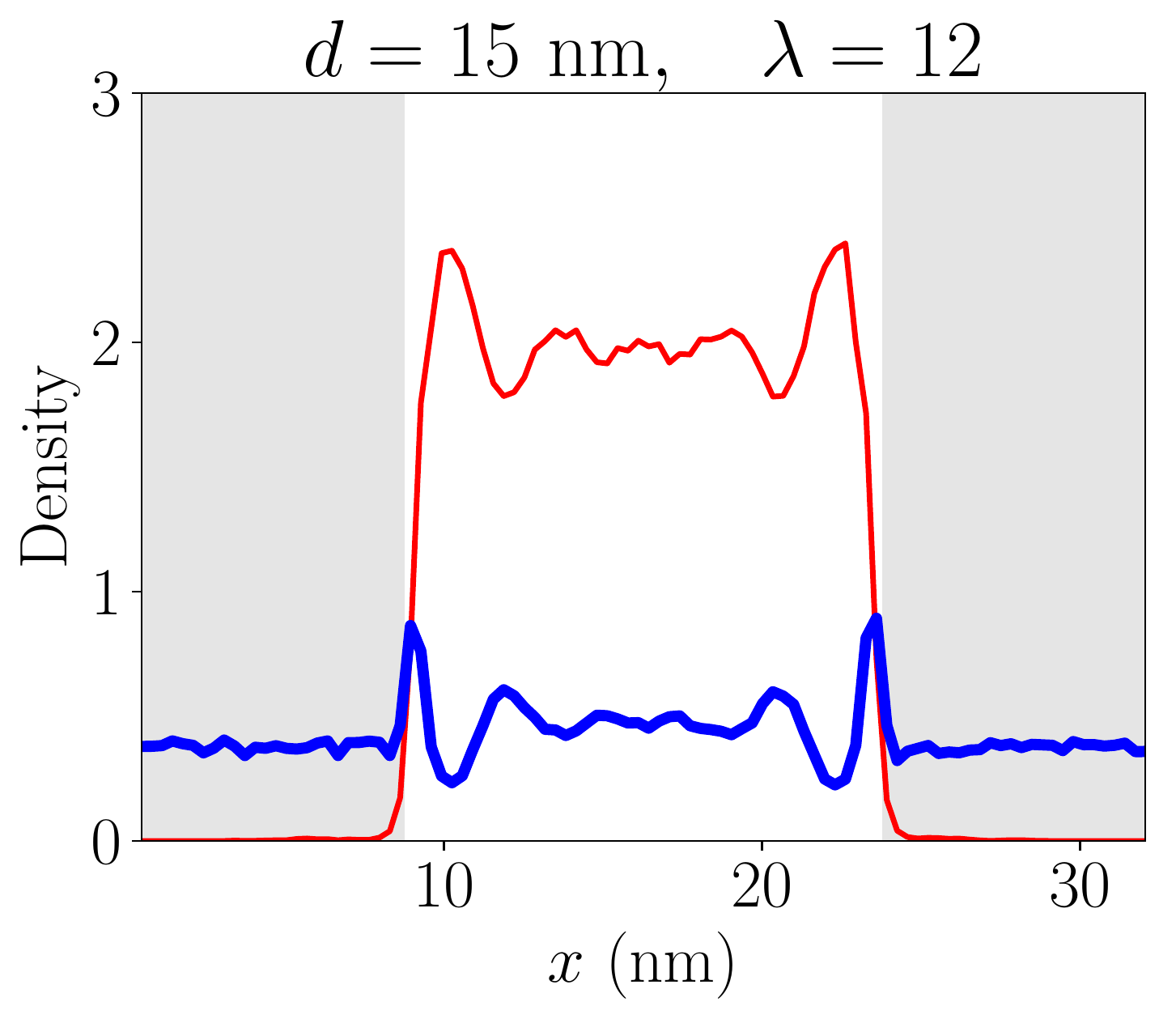}
~
\includegraphics[width=0.48\columnwidth]{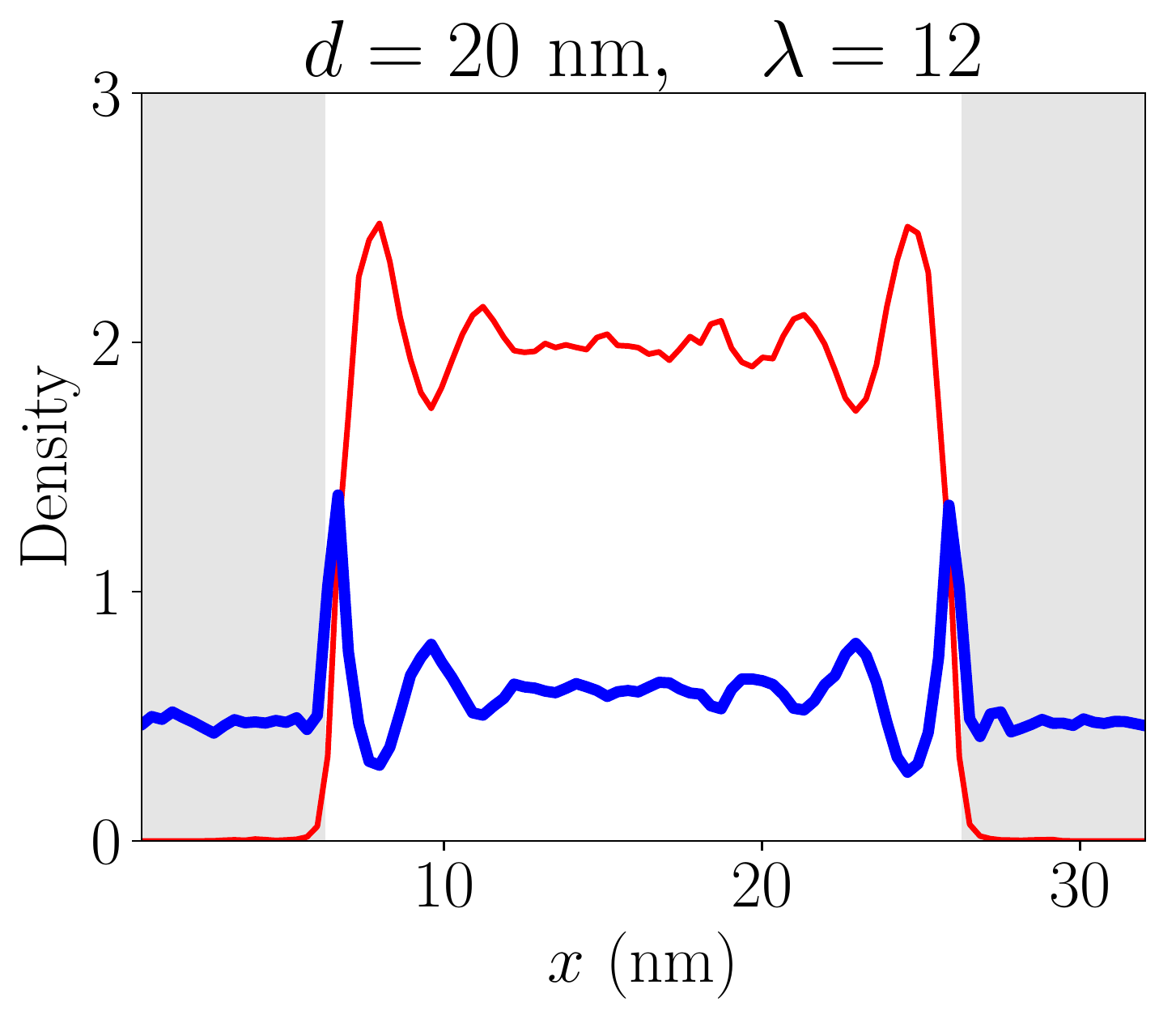}
\\
\includegraphics[width=0.48\columnwidth]{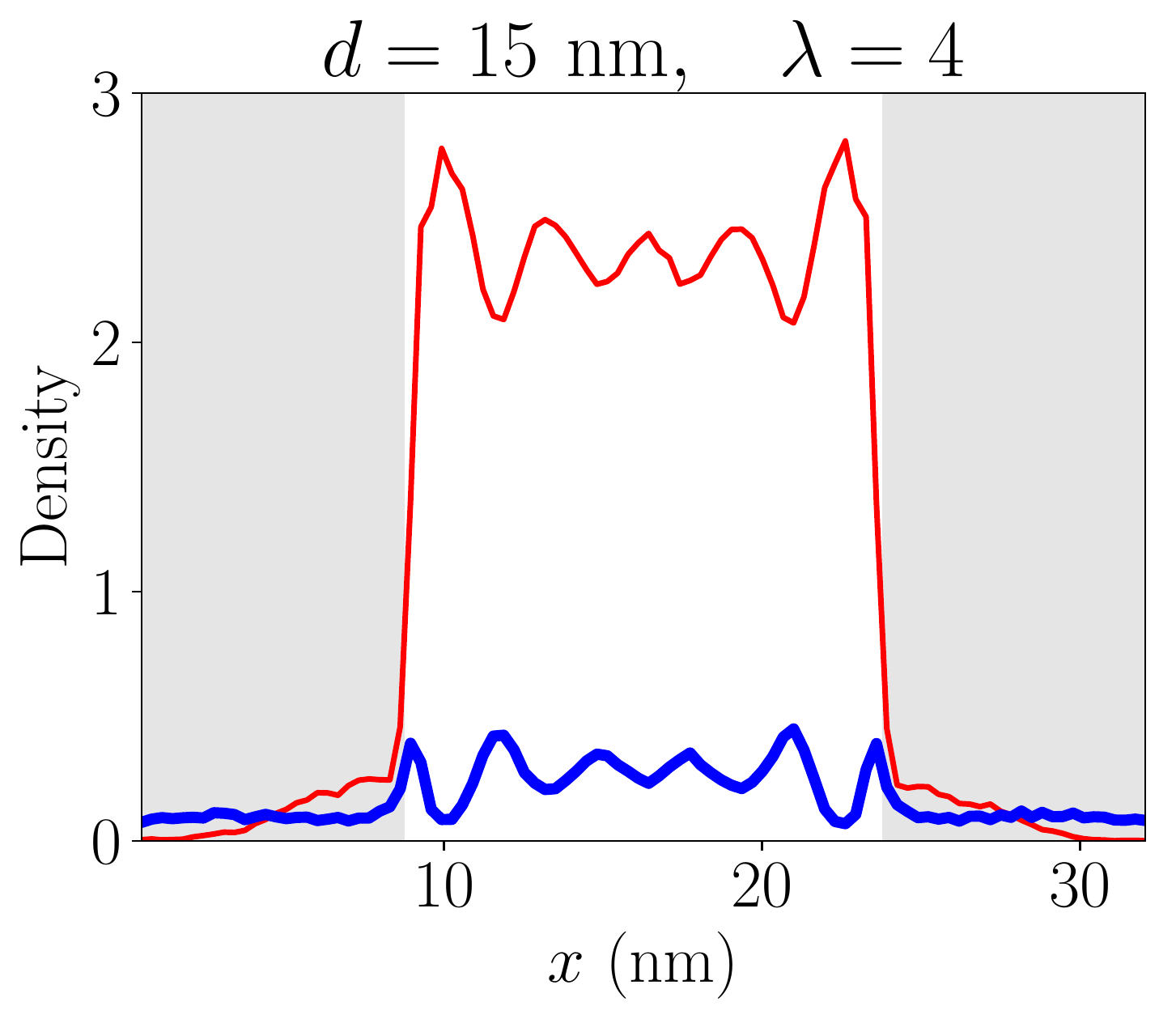}
~
\includegraphics[width=0.48\columnwidth]{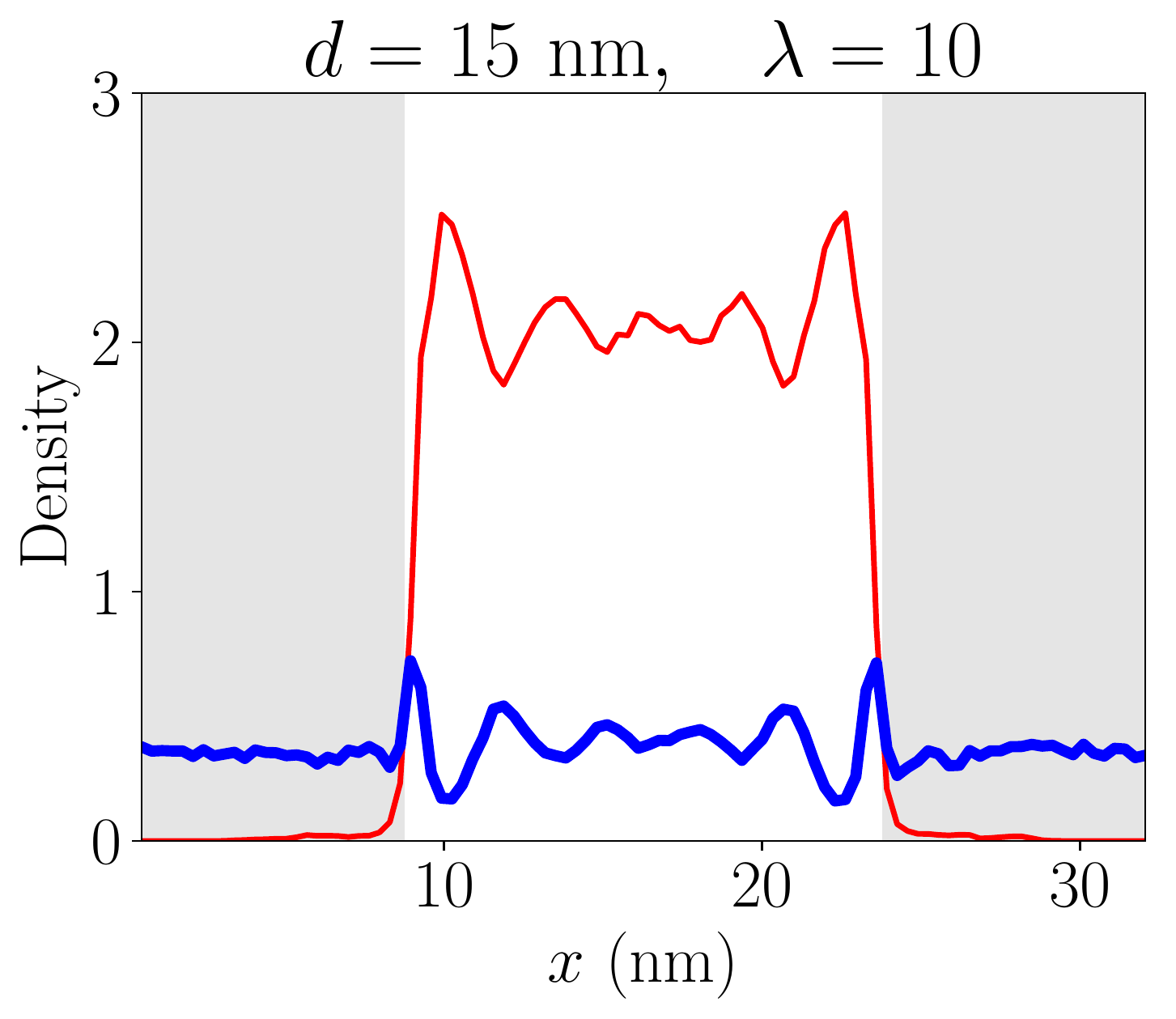}
~
\includegraphics[width=0.48\columnwidth]{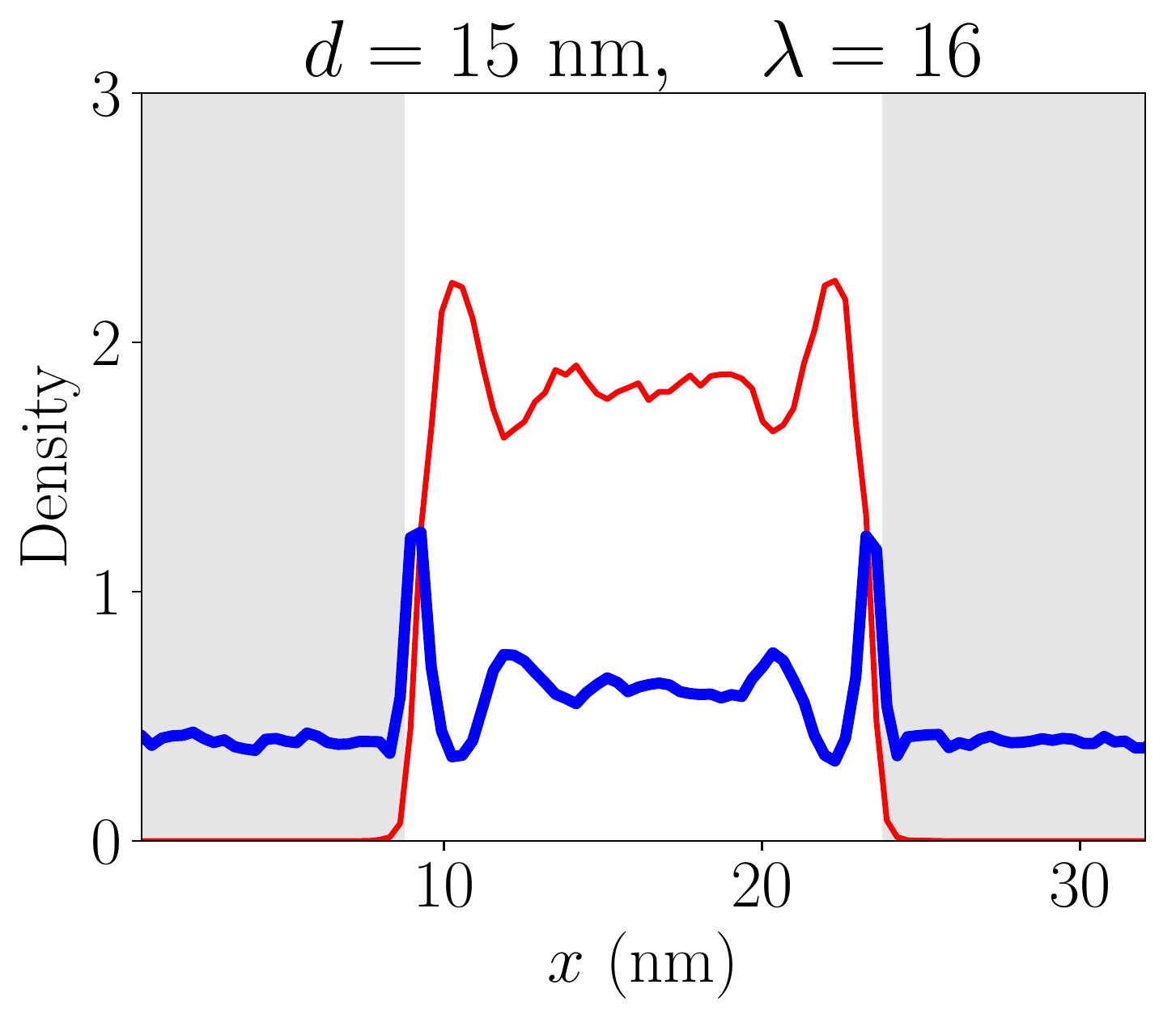}
~
\includegraphics[width=0.48\columnwidth]{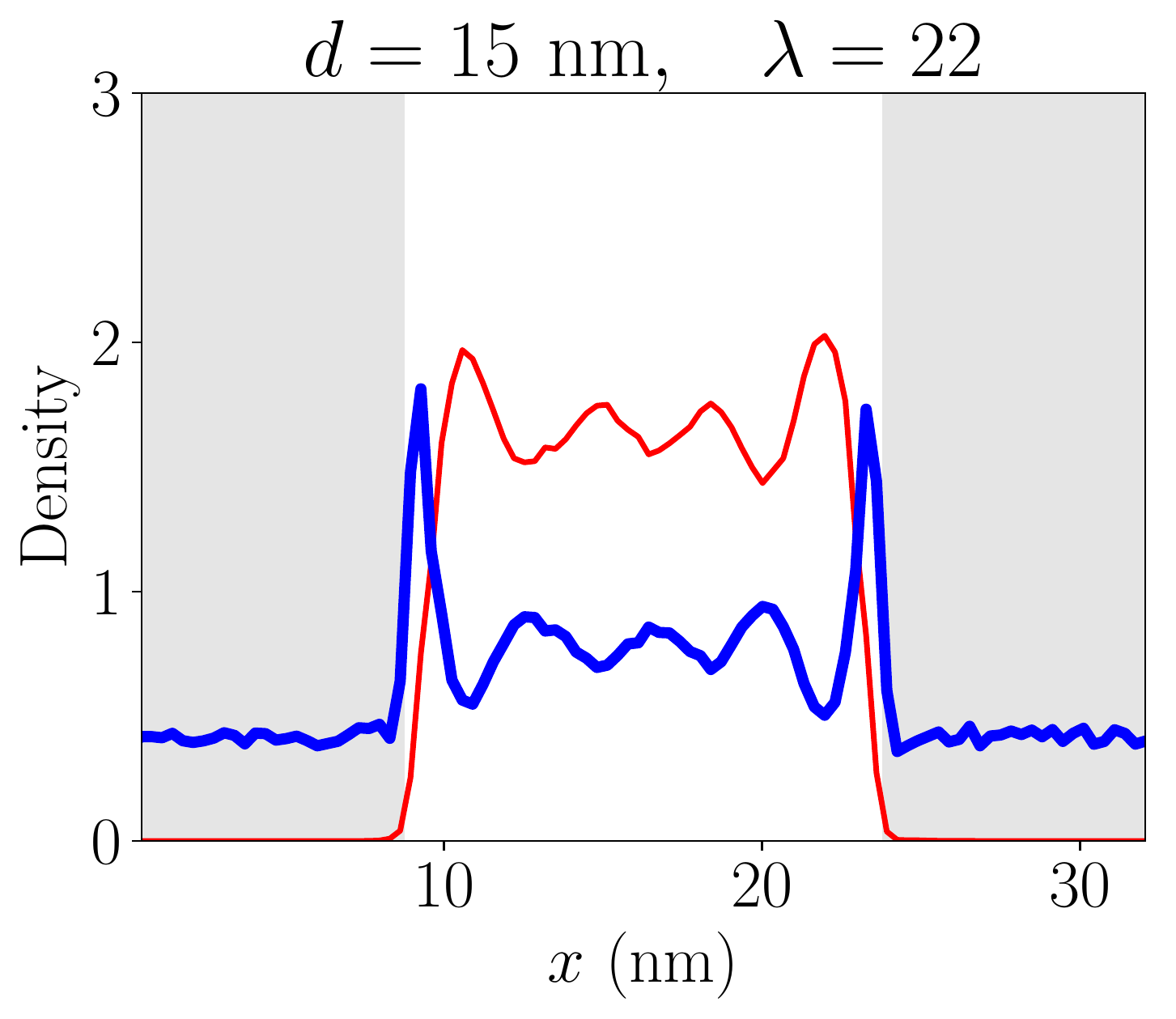}
\caption{(Color online) Top row: Profiles of water (blue) and PTFE backbone (red) confined by quartz for a range of film thicknesses $d = 5,10,15,20$~nm at water content $\lambda=16$.
Bottom row: Profiles of water (blue) and PTFE backbone (red) confined by quartz for a range of water contents $\lambda = 4,10,16,22$ at film thickness $d=15$~nm.}
\label{fig:profiles_quartz}
\end{figure*}

\subsection{Water distribution}
Our first step is to understand how water distribution in Nafion changes under the influence of confinement. Visual observation of the equilibrated boxes shows that clustering is established for carbon as well as quartz substrate (Fig.~\ref{fig:vmd} and supplementary material) for all water contents. For quartz, water is more dispersed in the backbone phase. A certain fraction of water has leaked into the confining material, which is expected due to the soft interparticle potential.

Despite the clustering created under the confinement of carbon, the size and structure of water clusters varies with the amount of water or film thickness. This is revealed by plotting the water distribution in the direction perpendicular to the Nafion film for carbon (Figs.~\ref{fig:profiles_carbon}) and quartz (Figs.~\ref{fig:profiles_quartz}). Here we observe different numbers of peaks in the water profile for each pair of $(d,\lambda)$. For low film thickness 5~nm, increasing water content creates a massive peak in the middle of the film. For larger thicknesses, notably 20~nm, the profile passes from as many as five peaks at $\lambda=4$ to one large peak at $\lambda=24$. We can deduce that at low $\lambda$'s water is clustered into peaks of typical size 4-5~nm, whereas for high $\lambda$'s most water concentrates in the middle of the film. All the profiles are shown in the supplementary material.

In contrast to carbon, hydrophilic quartz produces large peaks of water at the ionomer-substrate interface. Water peaks inside the ionomer are relatively smaller and keep their size with increasing amount of water, but their number changes with both film thickness and water content. For 20~nm, this number goes from five at $\lambda=4$ to three at $\lambda=24$.

Above $\lambda=16$, a distinct water depletion zone is formed at the ionomer-carbon interface, and a water saturation zone appears at the ionomer-quartz interface for all film thicknesses and water contents. Both these effects are an unambiguous sign of the hydrophobicity and hydrophilicity of carbon and quartz, respectively. 

\begin{figure*}
\centering
\includegraphics[width=0.63\columnwidth]{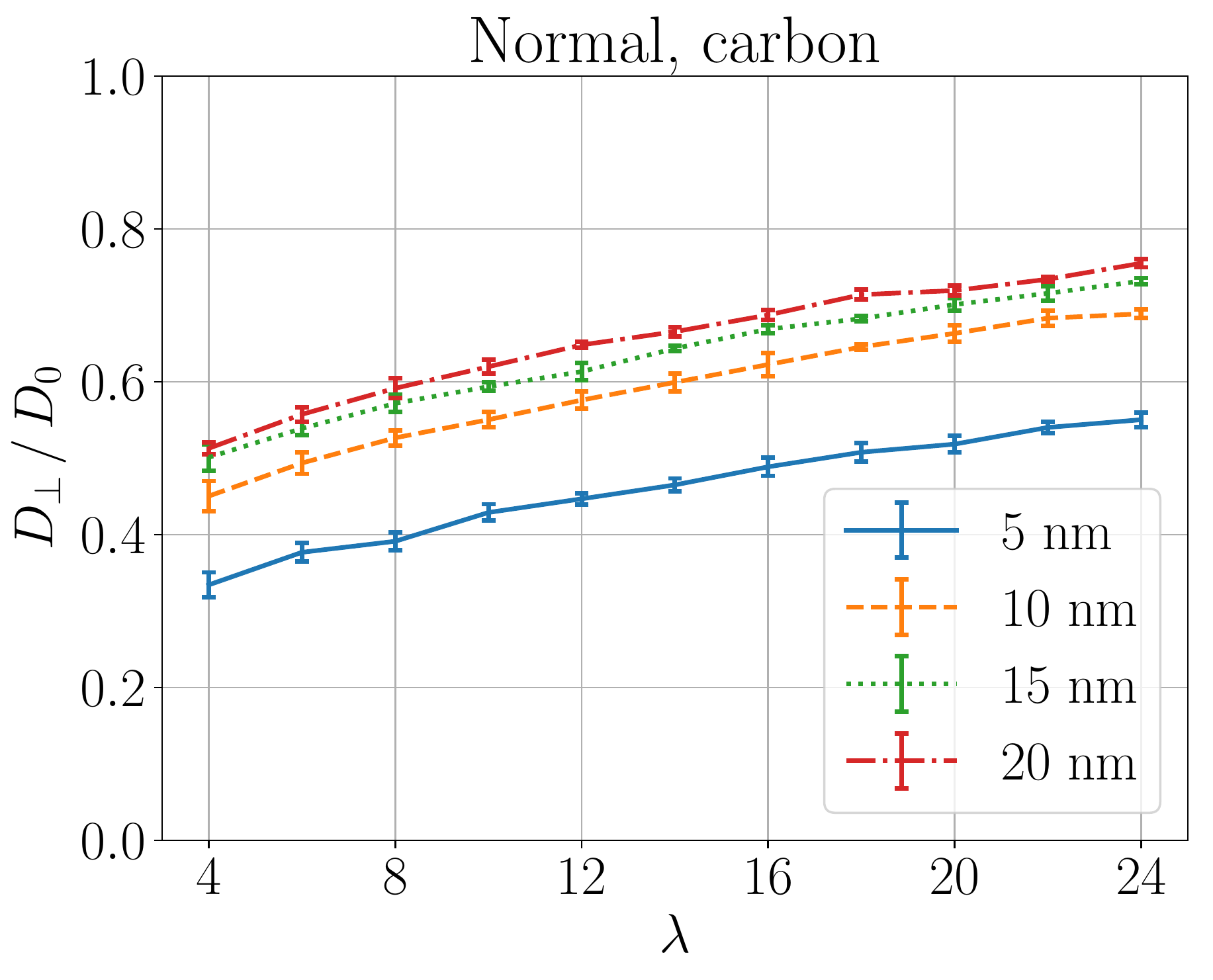}
~
\includegraphics[width=0.63\columnwidth]{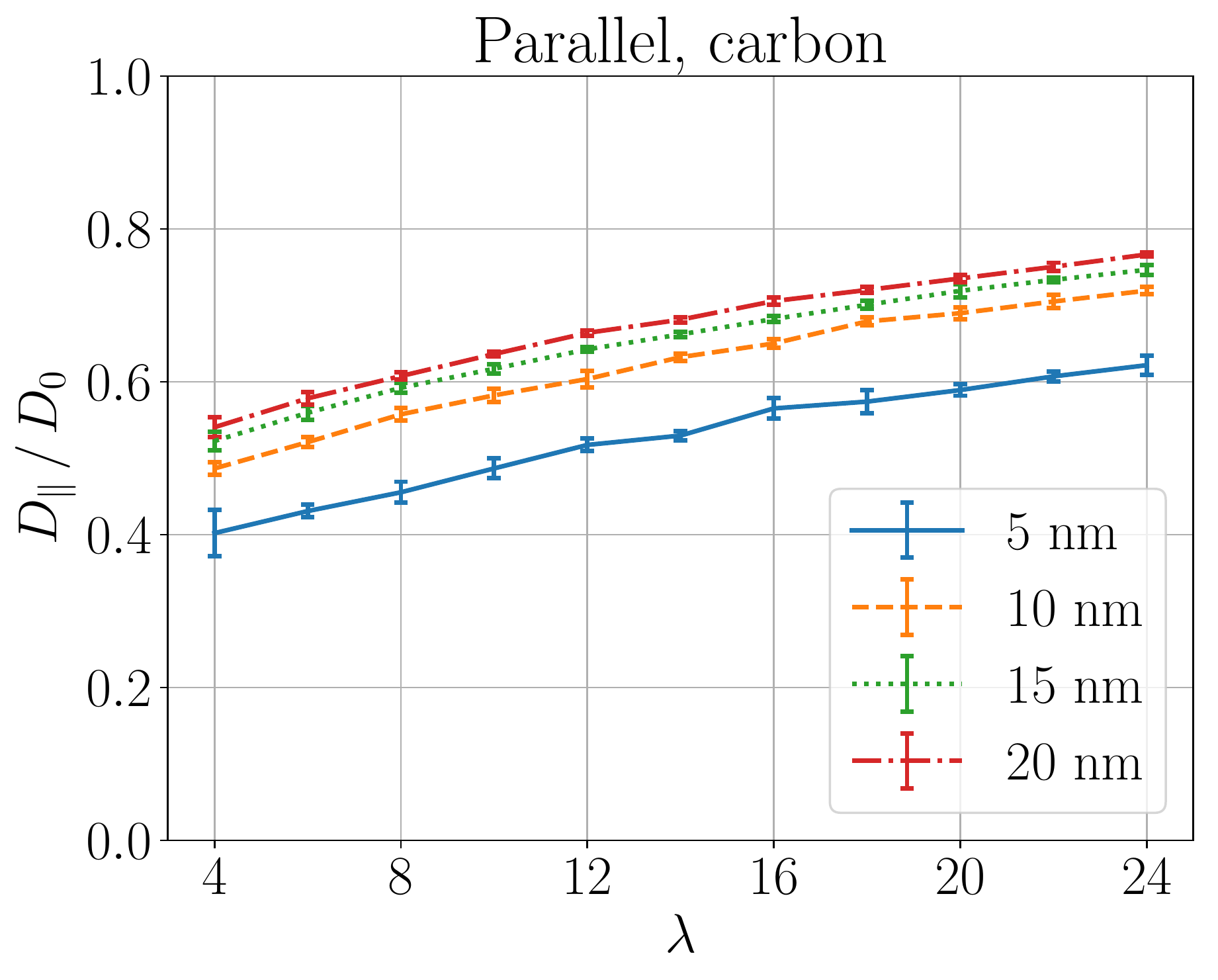}
~
\includegraphics[width=0.63\columnwidth]{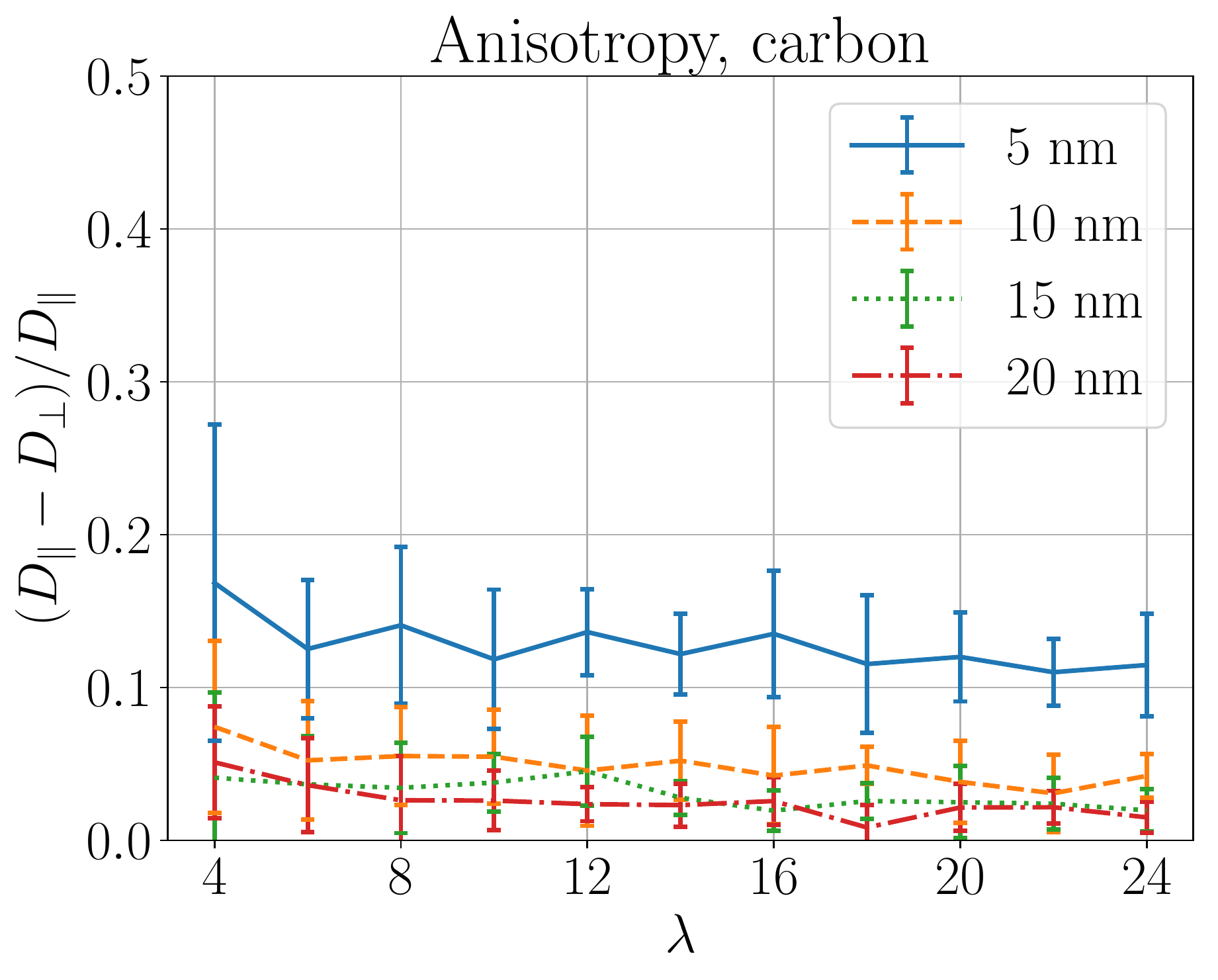}
\\
\includegraphics[width=0.63\columnwidth]{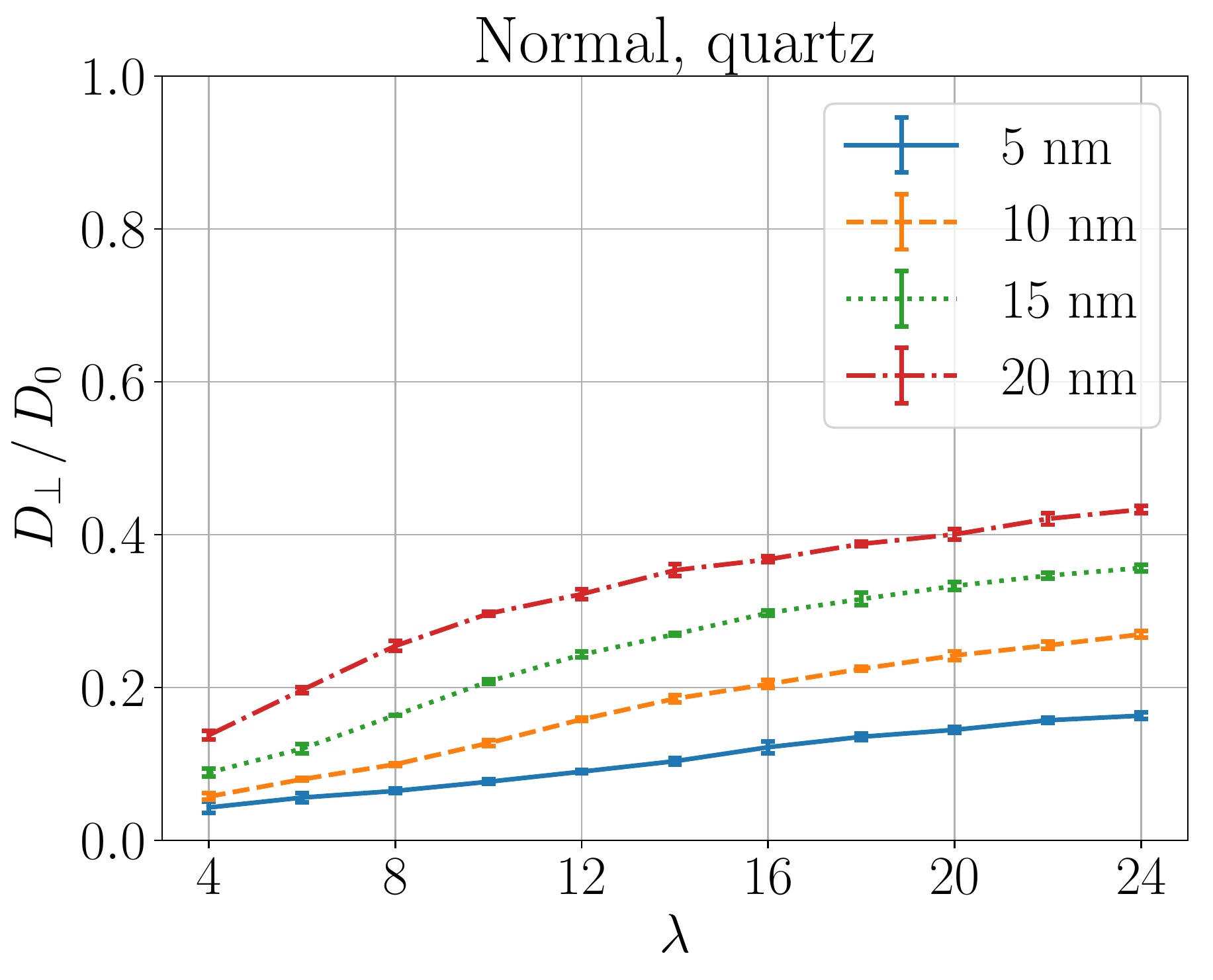}
~
\includegraphics[width=0.63\columnwidth]{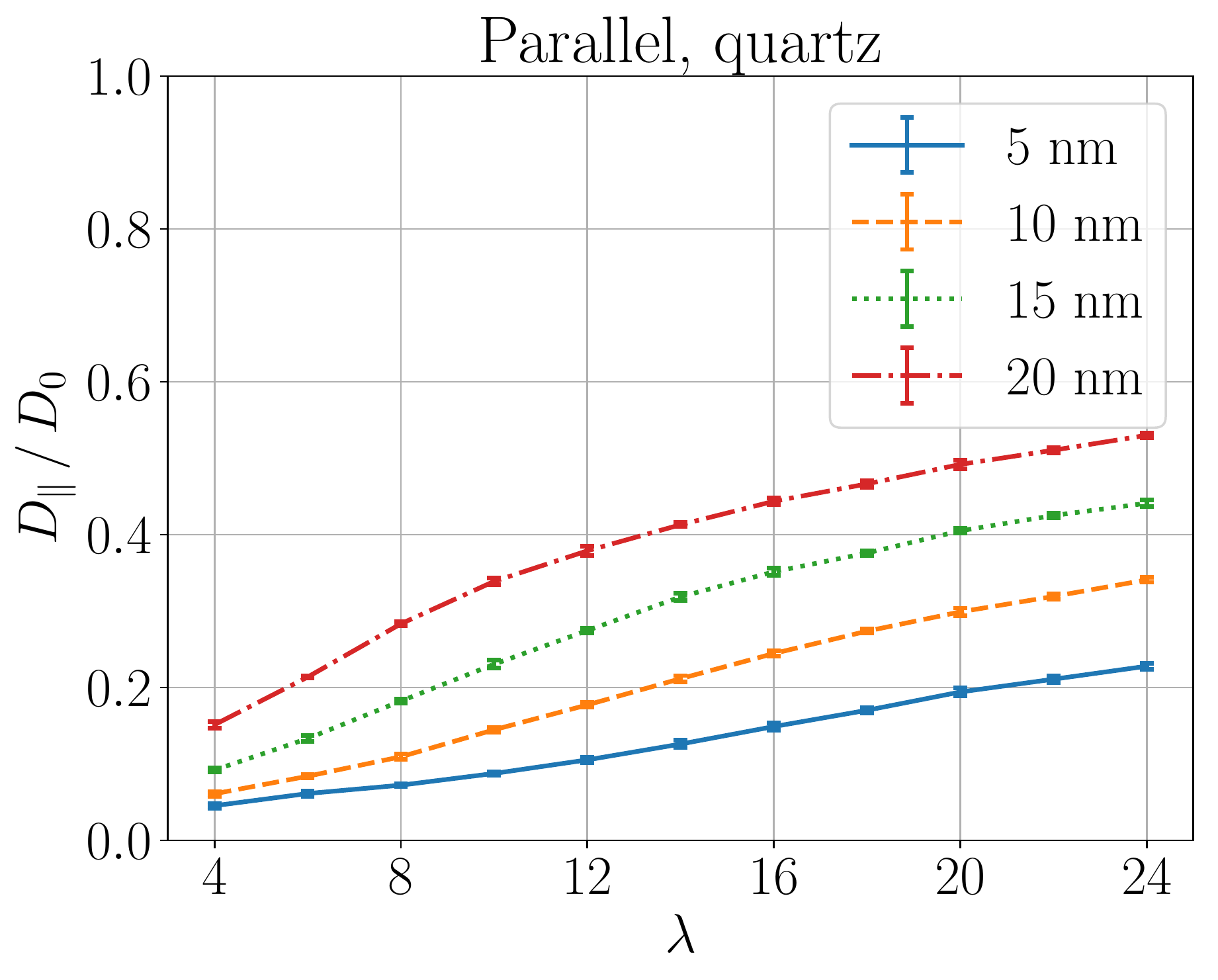}
~
\includegraphics[width=0.63\columnwidth]{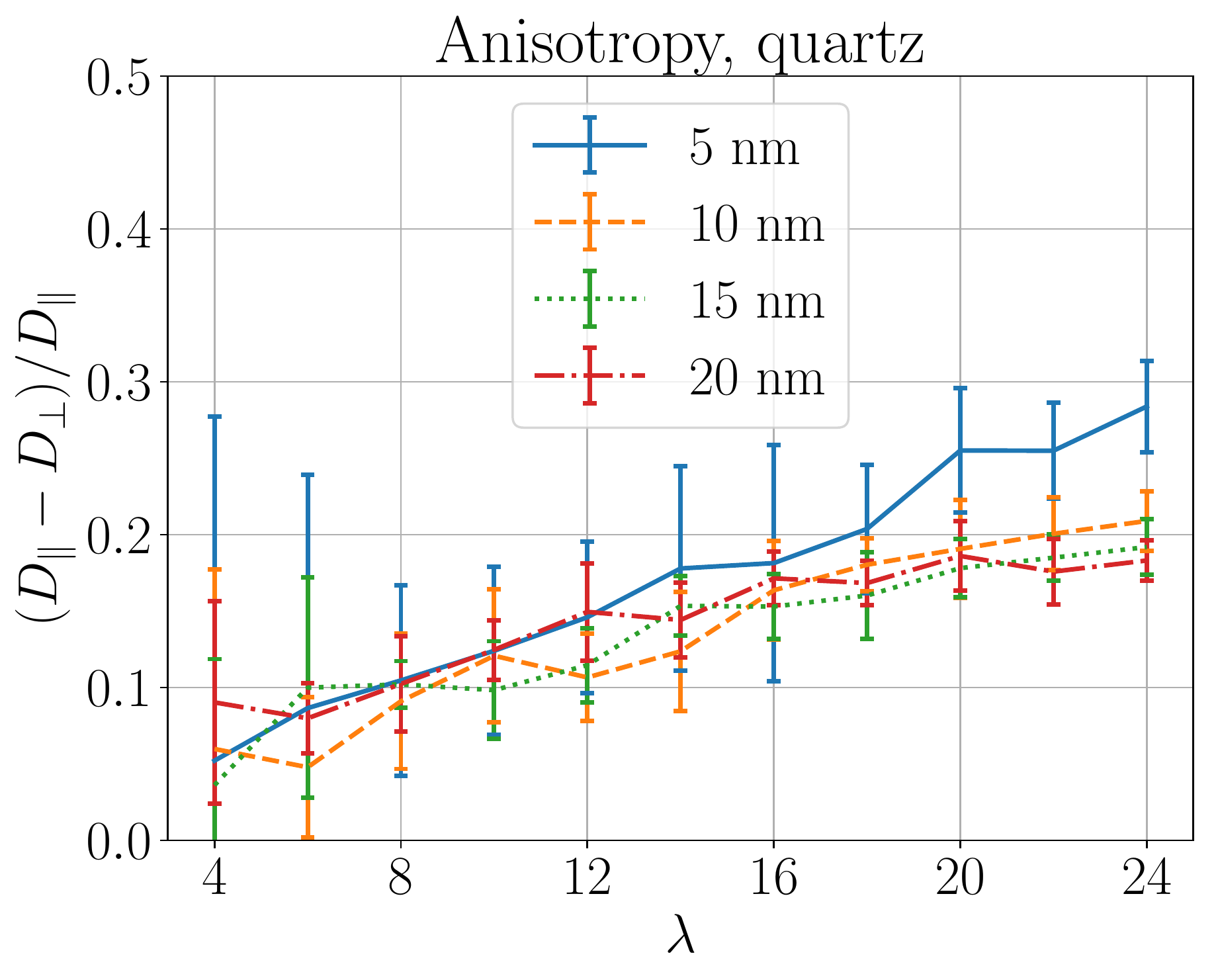}
\caption{(Color online) Parallel $D_\|$ and normal $D_\bot$ diffusivities of water for Nafion film confined by carbon (top row) and quartz (bottom row) as a fraction of the self-diffusion coefficient $D_0$ of a pure DPD liquid at $a=25\,\kT$. Right column: Percentage difference between parallel and normal diffusivity.}
\label{fig:diff}
\end{figure*}

\subsection{Diffusivity}
Having analysed the static properties of water in confined Nafion films we proceed with the dynamics. It would be too much to expect exact quantitative agreement from DPD, but the overall trends in behaviour are well captured.

To understand the dynamics of our system we calculated the self-diffusion coefficient of W beads as a function of film thickness and water content. Figs~\ref{fig:diff} (top) show curves of water self-diffusion w.r.t. $\lambda$ for multiple film thicknesses between carbon substrates. The gaps between these curves become narrower as the film thickness increases: the widest gap is between curves at 5 and 10~nm and the narrowest between 15 and 20~nm. More importantly, the asymmetry between normal and parallel diffusivities is most pronounced at 5~nm film, and gradually decreases as the thickness increases. The diffusivity of water between quartz substrates (Figs.~\ref{fig:diff}, bottom) shows different features: the curves of same film thickness are separated by the same distance at larger water contents, and the gap between parallel and perpendicular diffusivity narrows.

We define the measure $\eta=(D_{\|} - D_{\bot})/D_{\|}$ to quantify the anisotropy between parallel and normal diffusivities. Figs.~\ref{fig:diff} show strong dependence of $\eta$ on film thickness in case of carbon confinement. This effect is much smaller in case of quartz and is only present at high water contents $\lambda>20$. For carbon, there is weak depependence of $\eta$ on the water content, whereas in case of quartz $\eta$ rises linearly with $\lambda$. The error bars at $\lambda=4$ and 6 suggest insufficient sampling due to low number of water beads in the simulation box, but they do not prevent from deducing the overall trend.

We note that the diffusivity anisotropy of a film confined by quartz does not disappear with increasing film thickness, in sharp contrast with carbon. We attribute this phenomenon to the excess water at the film-quartz interfrace creating a layer in which the W beads can move relatively freely in parallel to the film. As the water profiles on Fig.~\ref{fig:profiles_quartz} demonstrate, this layer becomes more pronounced with rising water content and irrespective of the film width, which also explains the steady increase of $\eta$.

These findings suggest explanations for behaviour in the catalyst layer of fuel cells. Water freshly formed from protons and oxygen on the surface of platinum nanoparticles has a number of transport routes: it can either remain within the ionomer phase and move directly into the membrane (water back-diffusion); it can pass through the ionomer film covering the Pt and emerge as liquid water within the pore space of the catalyst layer, or it can evaporate from the surface of the ionomer film into the pore space. The balance of these pathways will depend on the operating conditions of the cell (relative humidity, current density, pressure) and the position of the Pt within the catalyst layer in x, y and z directions. Clearly, the water content within, under and on top of the ionomer film will have a significant impact on the transport of oxygen to the Pt surface and it may not be necessary to suppose that the thin film of ionomer has inherently lower oxygen permeability as others have done;~\cite{Jinnouchi_EA_2016,Kongkanand_JPCL_2016} or it may be that both water build up and low ionomer permeability, caused by unusual structuring of the thin films, account for the local oxygen transport issue. In this context however, simulations of thin Nafion films on platinum surfaces would be highly desirable to clarify which of these explanations is more likely.

\subsection{Water connectivity}
The diffusivity of water is one of the ways to describe general transport properties. But this quantity is insufficient if the particles move around by quantum tunnelling, like protons in ionomer membranes do. Mesoscale methods cannot directly capture protonic conductivity, as there are no free protons, but the percolation of water network in the membrane is a suitable proxy. Hsu \emph{et al.} argued that percolation can explain the insulator-to-conductor transition in Nafion,~\cite{Hsu_MA_1980} and showed that the conductivity $\sigma$ satisfies a simple power law: $\sigma = \sigma_0(\lambda-\lambda_{\text{c}})^s$, where $\lambda$ is the water content~\cite{note_wc} and $s$ the critical exponent.

According to the percolation theory, the same scaling applies to the percolation cluster strength.~\cite{Stauffer_book_2003, Kirkpatrick_RMP_1973} Assuming a lattice in two or more dimensions and filling the sites with probability $\lambda$, a macroscopic \emph{percolation cluster} spanning the whole lattice starts appearing for $\lambda>\lambda_{\text{c}}$ and its size, the percolation cluster strength $P_\infty$, defined as the ratio of sites belonging to this cluster to the overall size of the lattice, grows as a power law:
\begin{equation}
P_\infty \sim (\lambda - \lambda_{\text{c}})^s,
\end{equation}
where $s$ is between 0.3 and 0.4 regardless of the lattice.~\cite{Kirkpatrick_RMP_1973}

We have used the ideas from percolation theory to understand the trends in the protonic conductivity in confined Nafion. We generated a water density map on a grid and set the cutoff for which a grid site still contains some water at 0.3, which is one tenth of the natural DPD density used in simulations. We then employed the flood fill algorithm to count the size of thus formed water clusters and observe how the largest one, the percolating cluster, varies with film thickness and water content.  

We evaluated two site percolations: on a two-dimensional grid formed from density profile of a slice through the middle of the thin film, and on a three-dimensional grid spanning the whole simulation box. 2D percolation channels and clusters can be easily visualised and so offer more intuition about the effect of the confining material on the thin film; 3D percolation should be a suitable approximation for the channels through which the protons move, based on the assumption that the protons will follow the best-hydrated pathways. The nominal value of the percolation cluster strength scales with the box size and is therefore not a good measure to compare across various film thicknesses. Therefore, we rescaled this value by the ratio of film thickness and the box size.

\begin{figure}
\centering
\includegraphics[width=0.22\columnwidth]{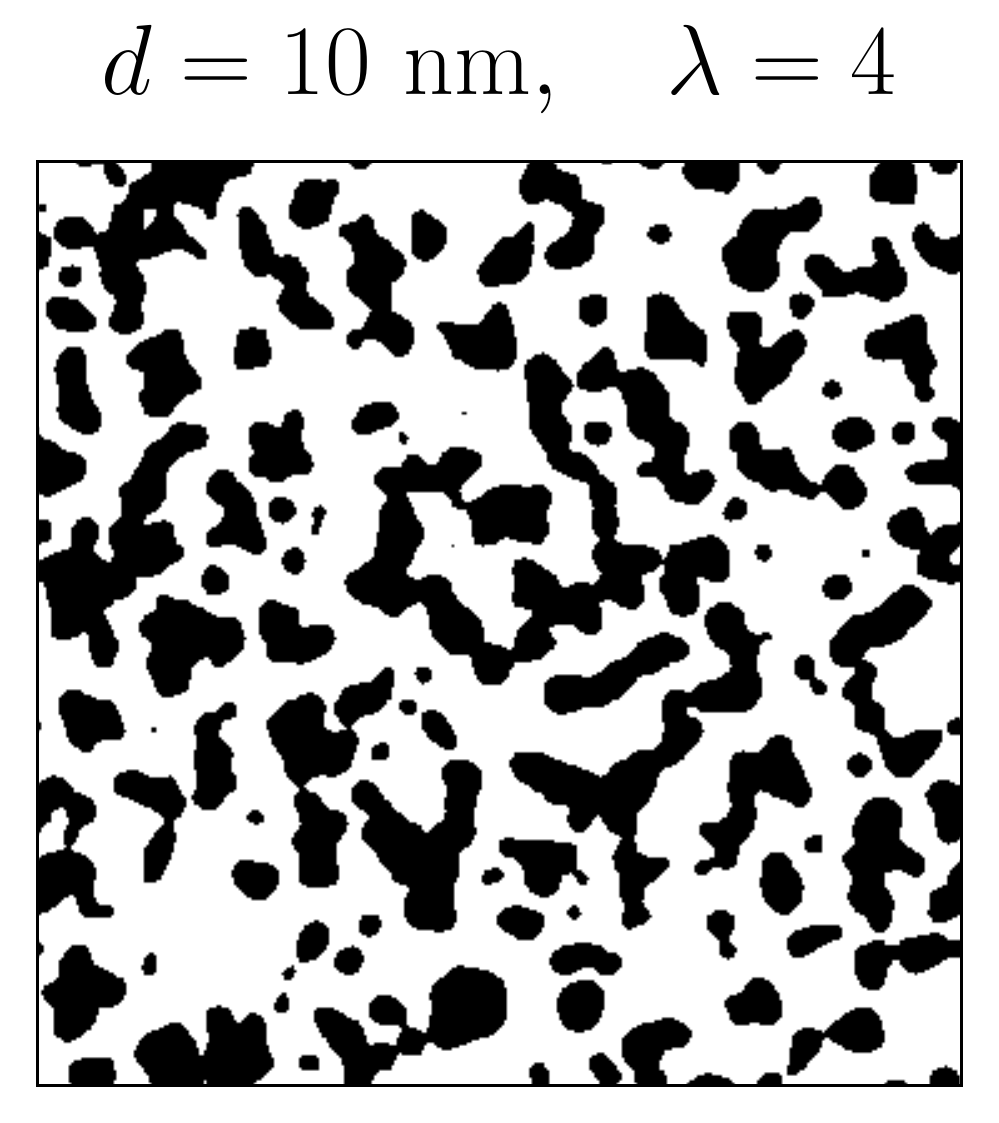}
~
\includegraphics[width=0.22\columnwidth]{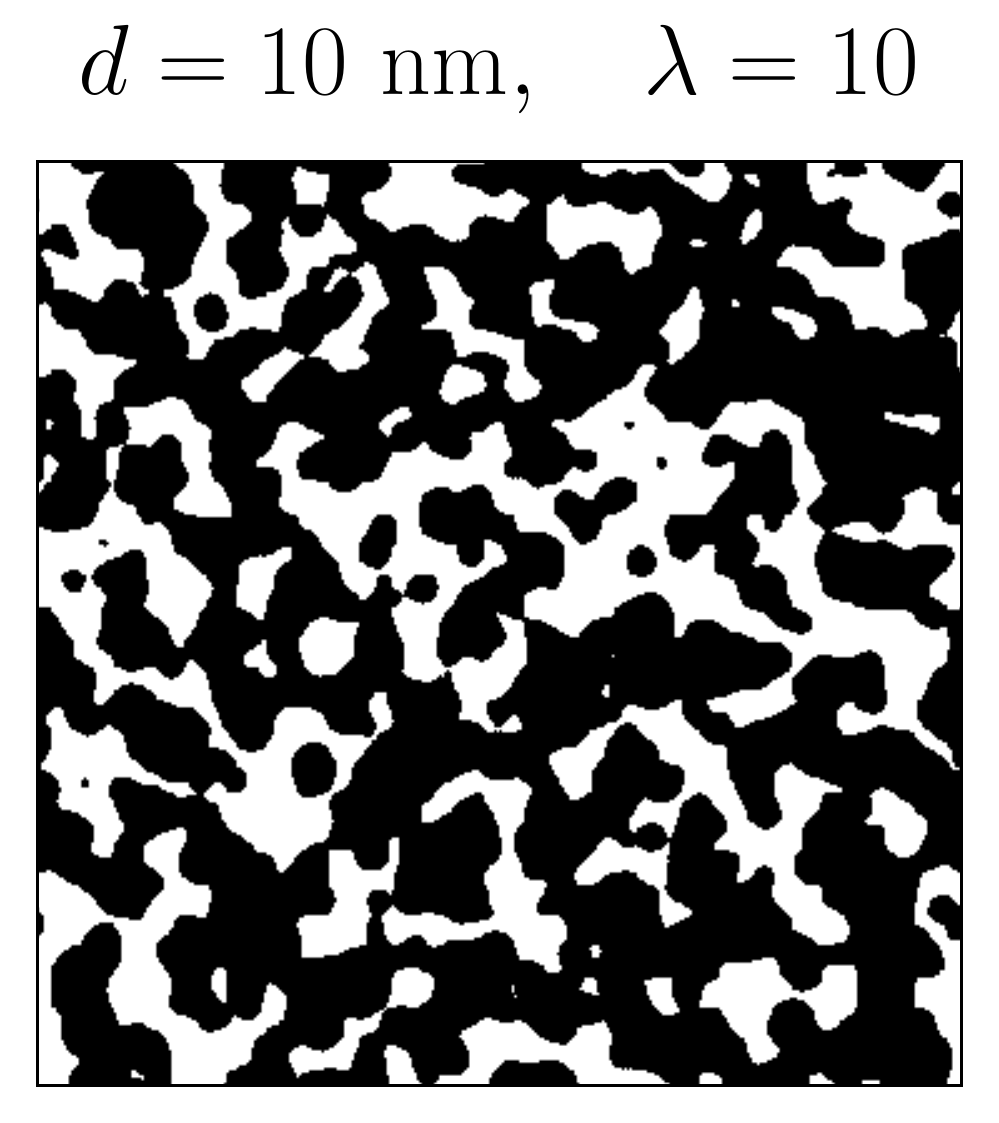}
~
\includegraphics[width=0.22\columnwidth]{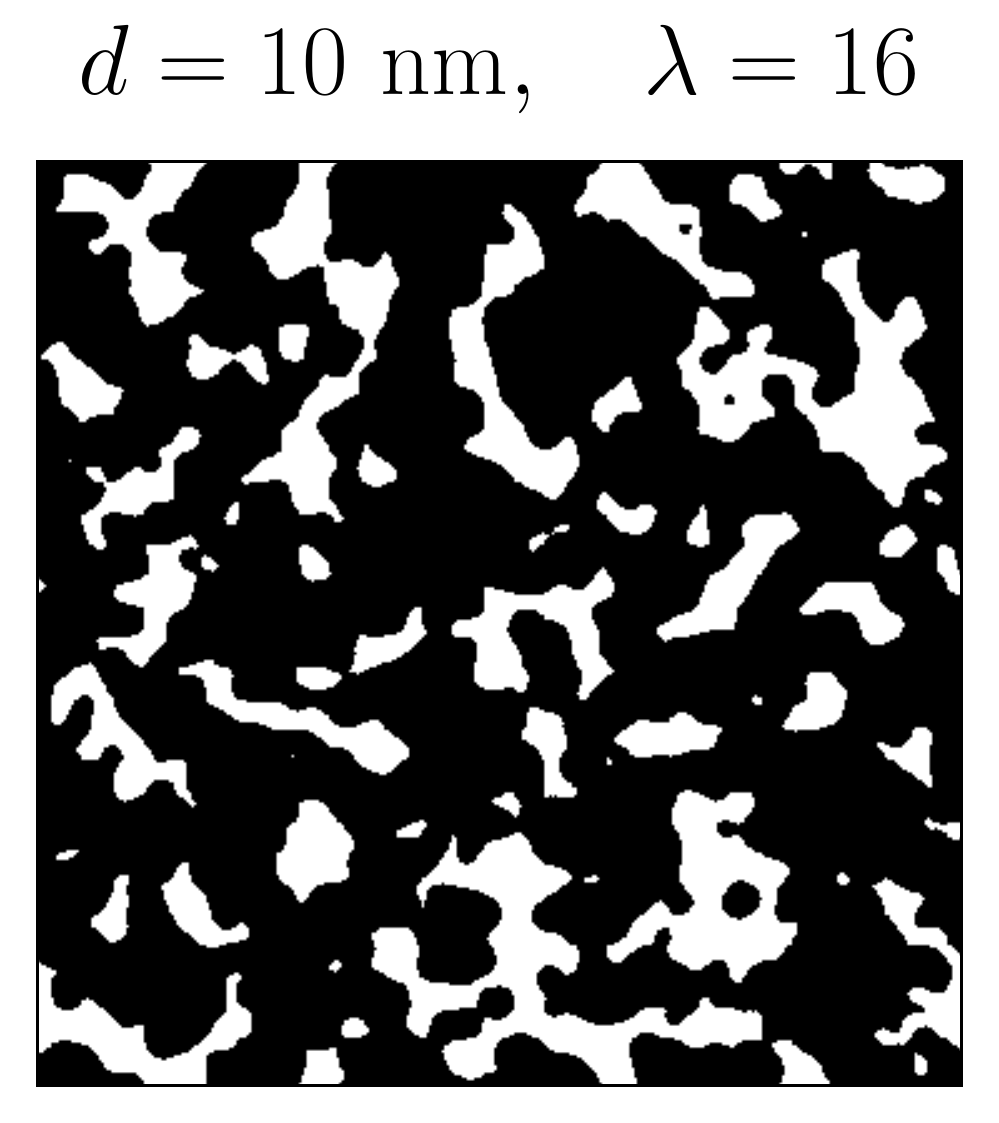}
~
\includegraphics[width=0.22\columnwidth]{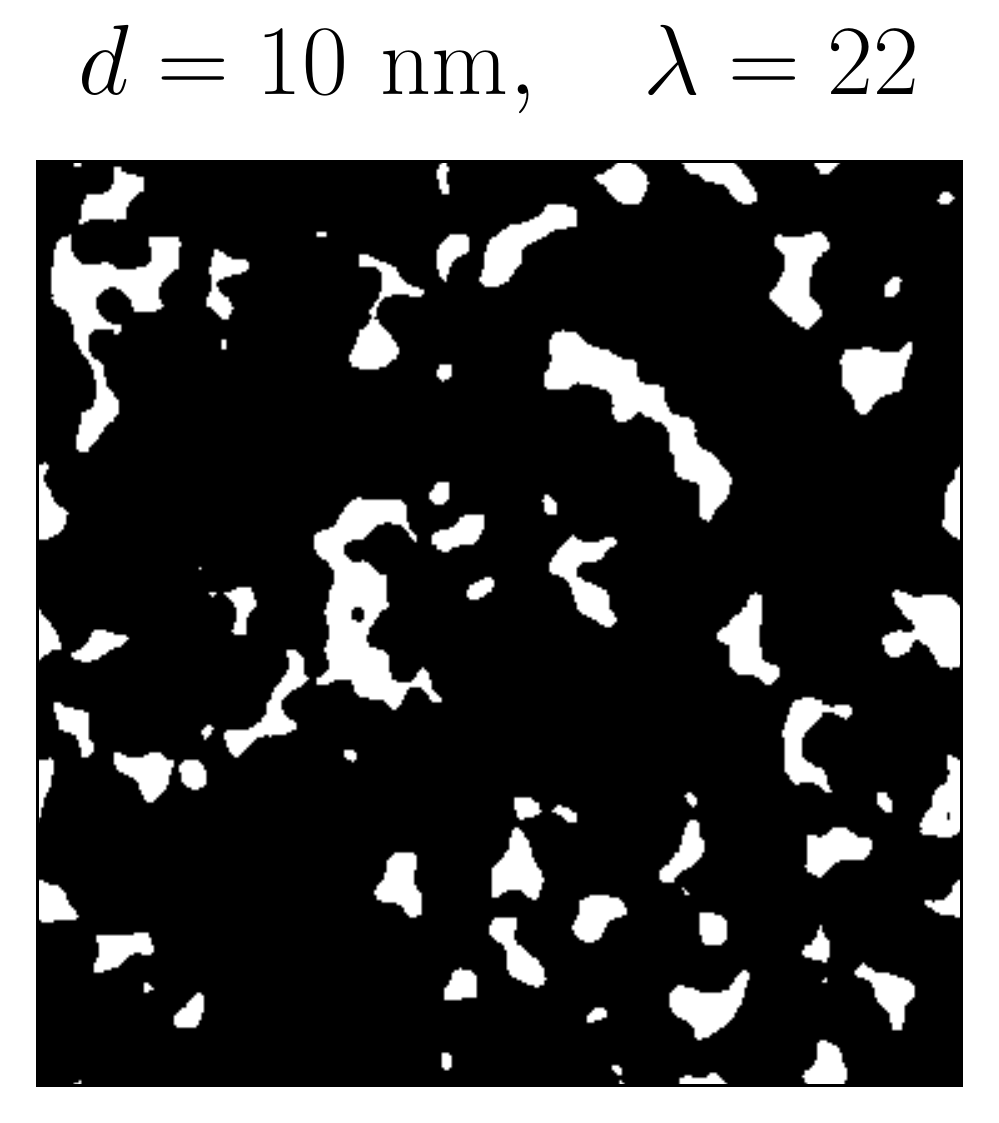}
\caption{Water clusters in thin film confined by carbon for film thickness 10~nm and water contents $\lambda = 4, 10, 16, 22$ respectively.}
\label{fig:carbon_clusters}
\end{figure}

\begin{figure}
\centering
\includegraphics[width=0.22\columnwidth]{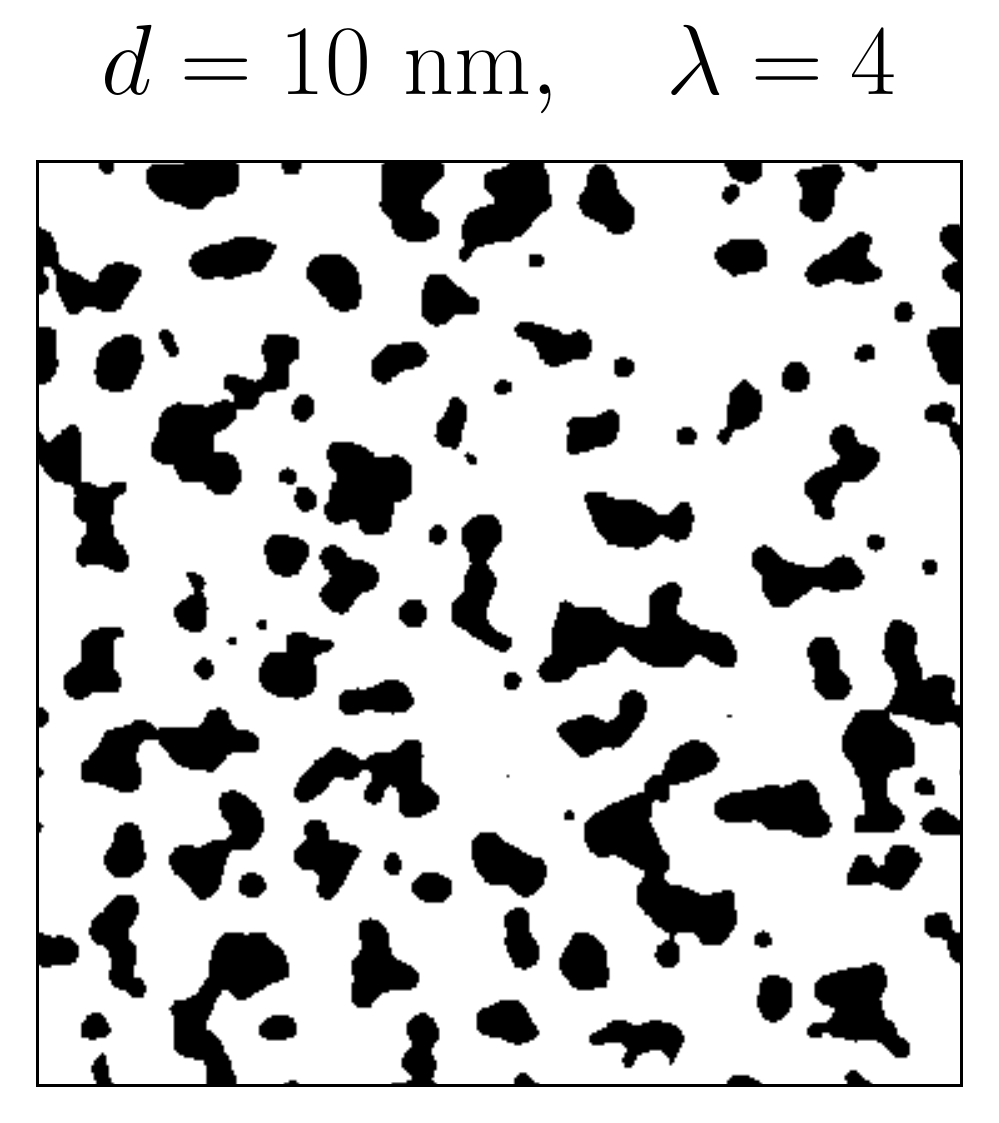}
~
\includegraphics[width=0.22\columnwidth]{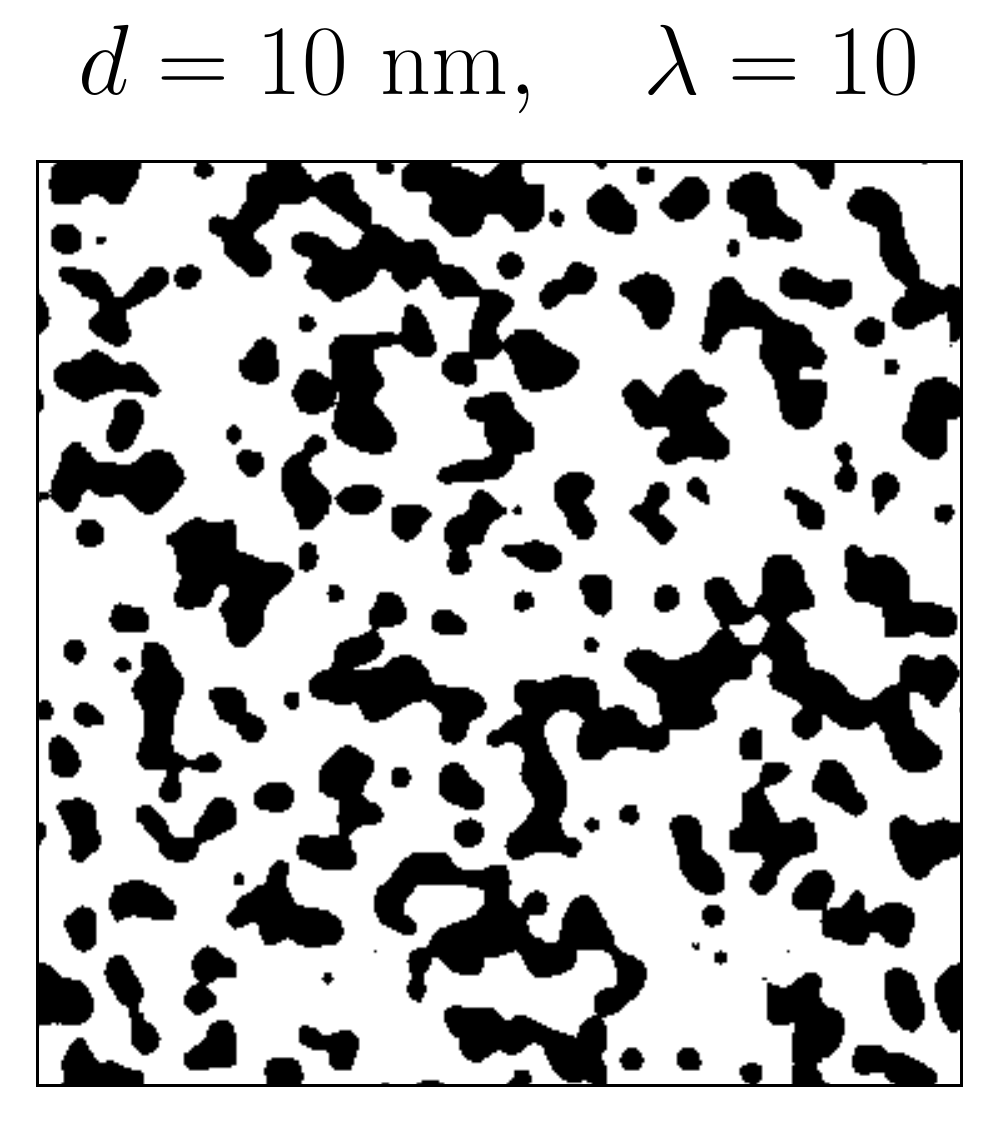}
~
\includegraphics[width=0.22\columnwidth]{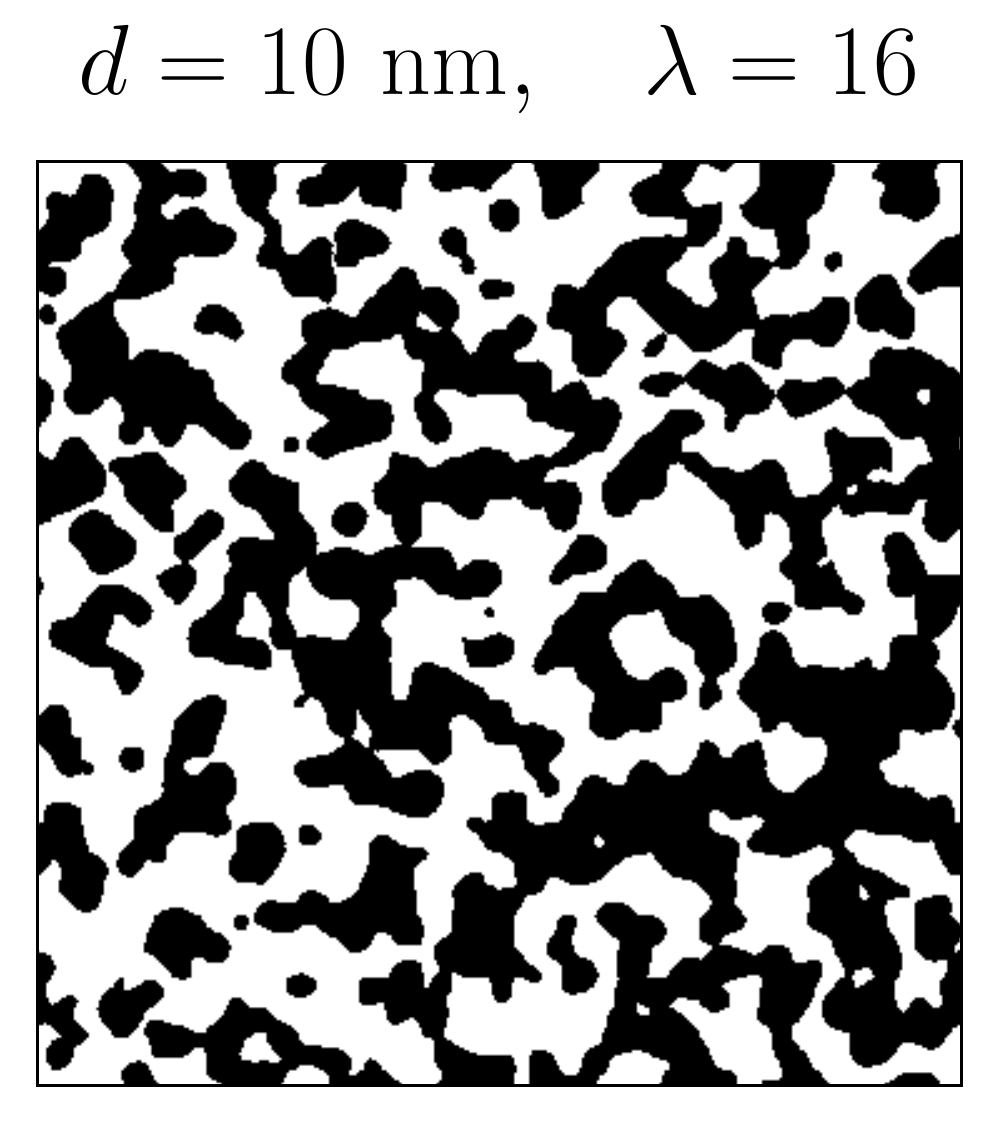}
~
\includegraphics[width=0.22\columnwidth]{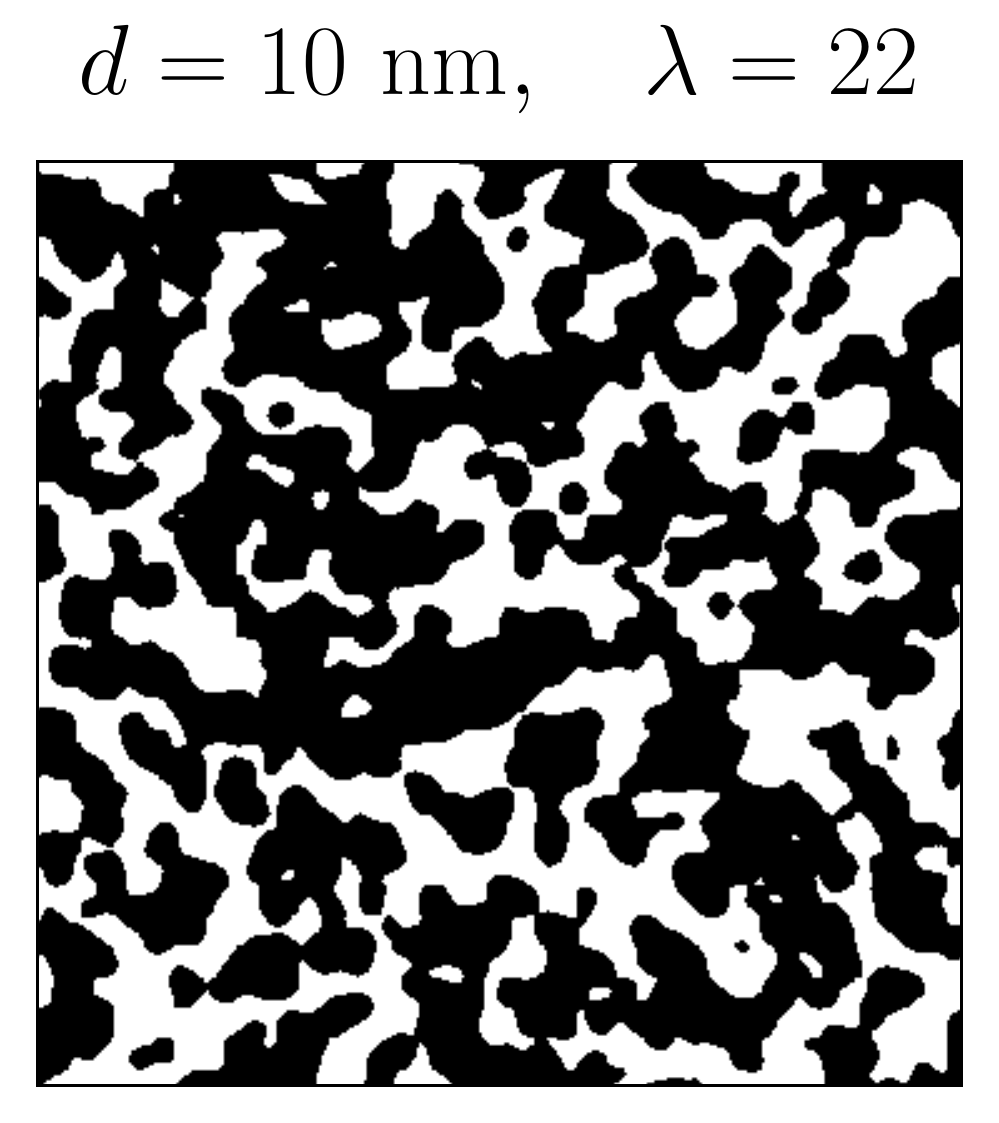}
\caption{Water clusters in thin film confined by quartz for film thickness 10~nm and water contents $\lambda = 4, 10, 16, 22$ respectively.}
\label{fig:quartz_clusters}
\end{figure}

Figs.~\ref{fig:carbon_clusters} and~\ref{fig:quartz_clusters} show 2D water clusters of a slice through a thin ionomer film confined by carbon and quartz, respectively (more can be viewed in the supplementary material). 

\begin{figure*}
\centering
\includegraphics[width=0.63\columnwidth]{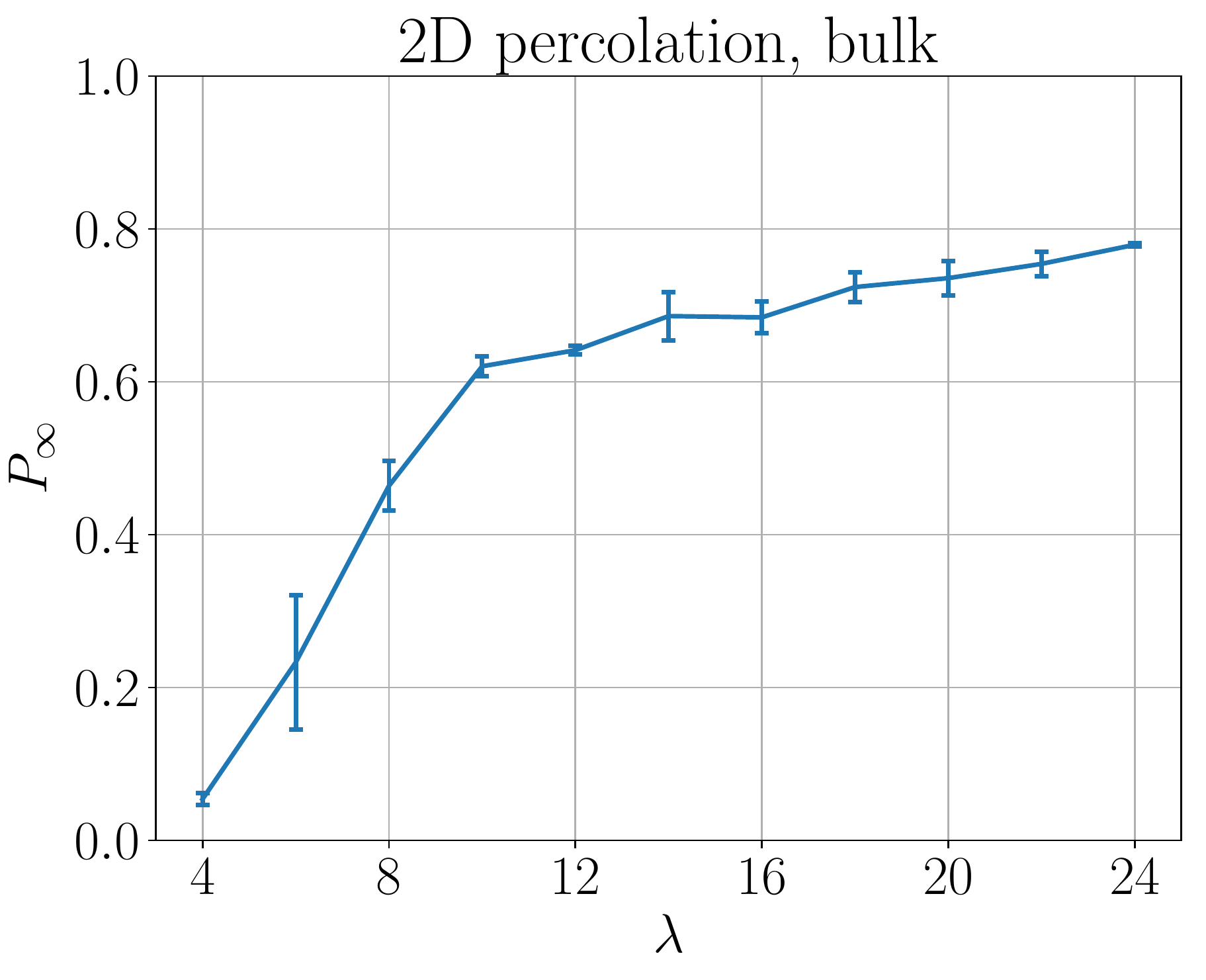}
~
\includegraphics[width=0.63\columnwidth]{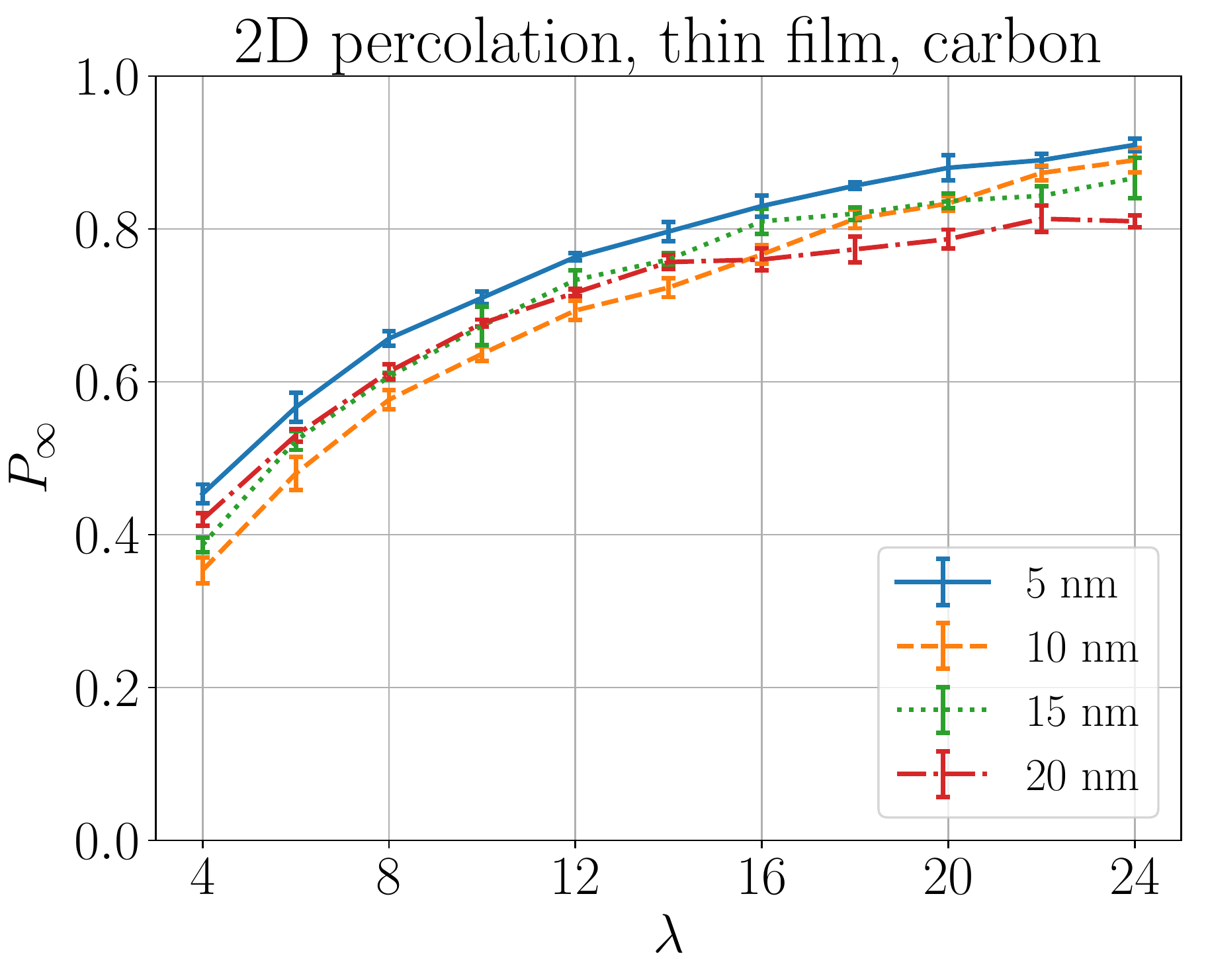}
~
\includegraphics[width=0.63\columnwidth]{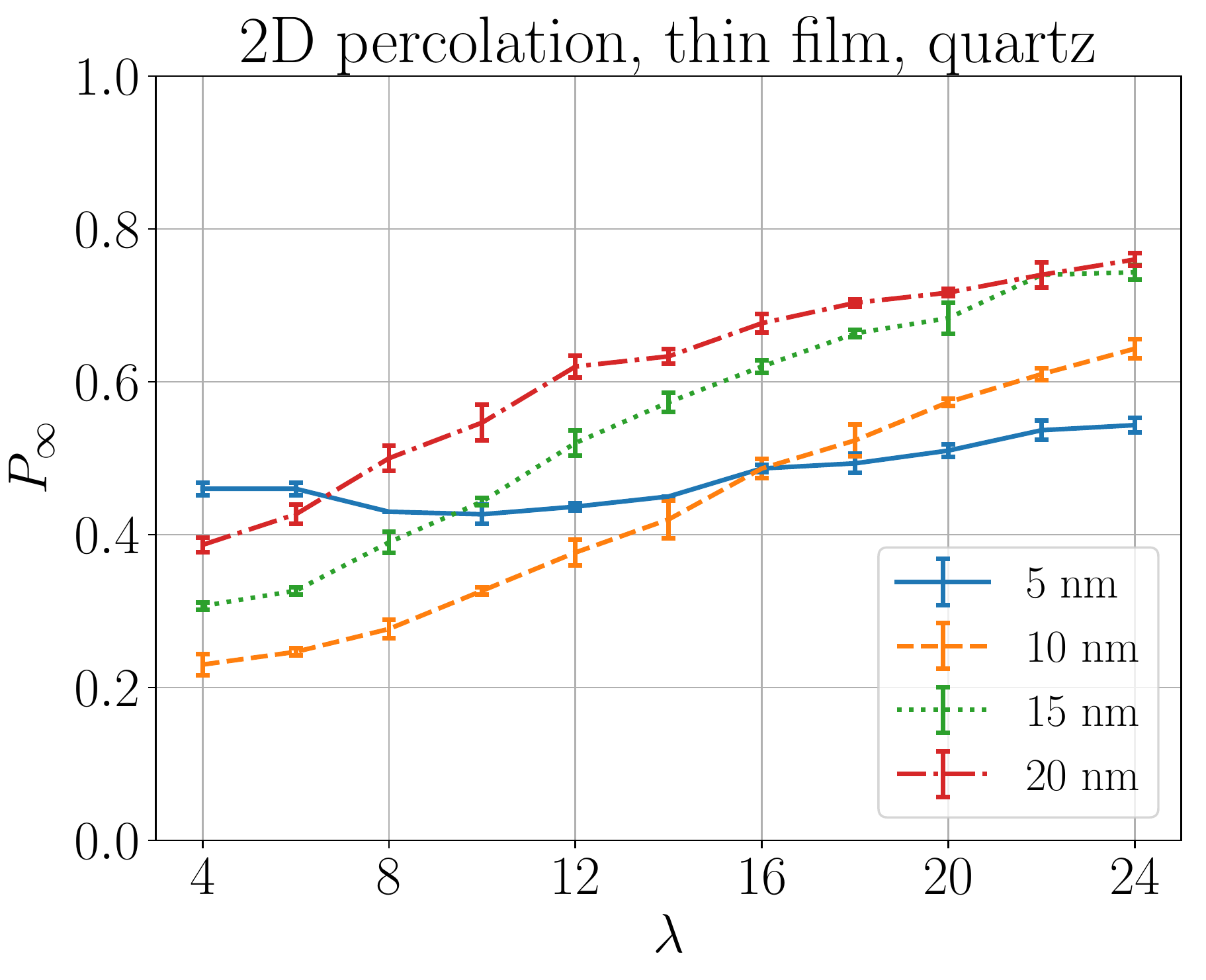}
\\
\includegraphics[width=0.63\columnwidth]{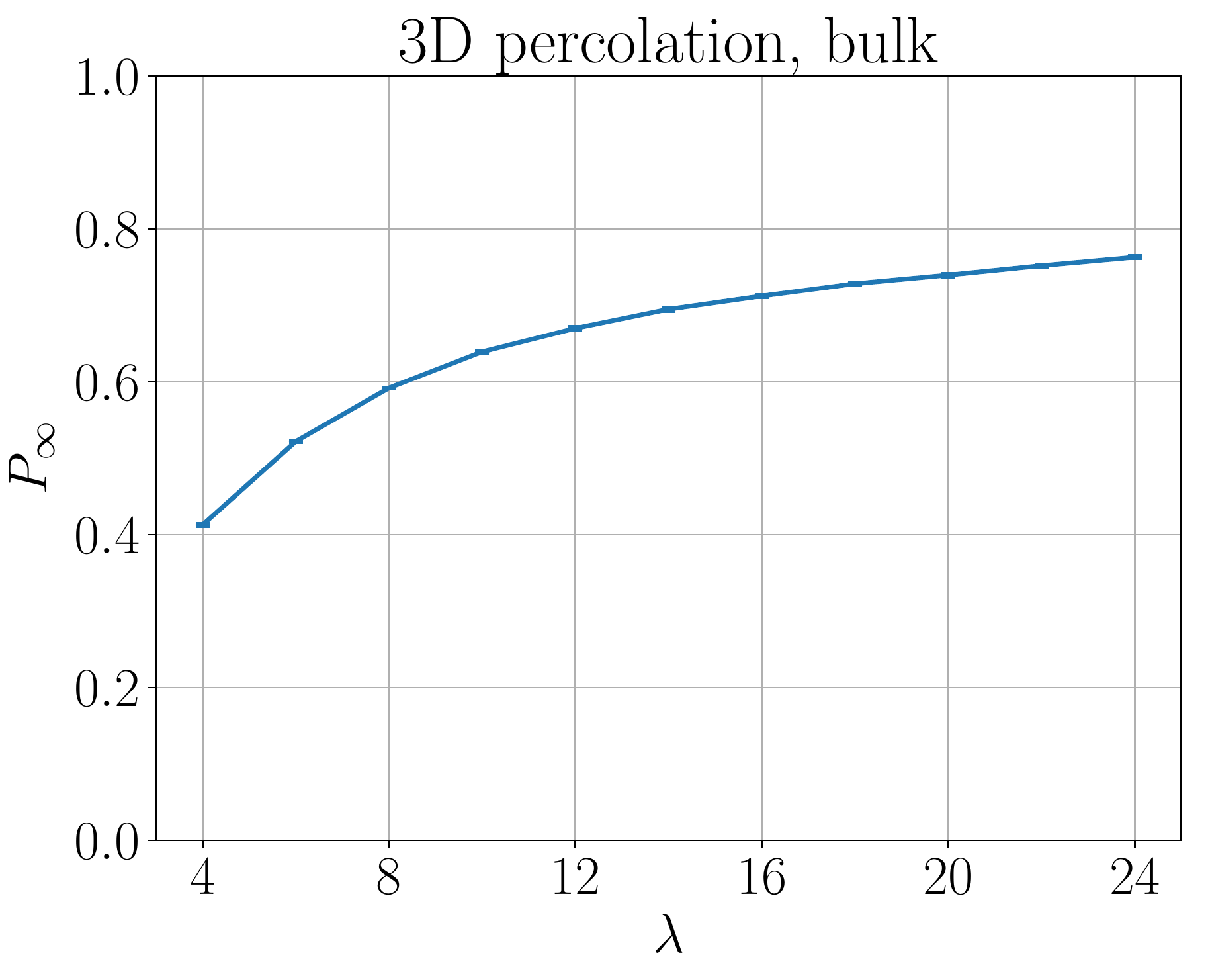}
~
\includegraphics[width=0.63\columnwidth]{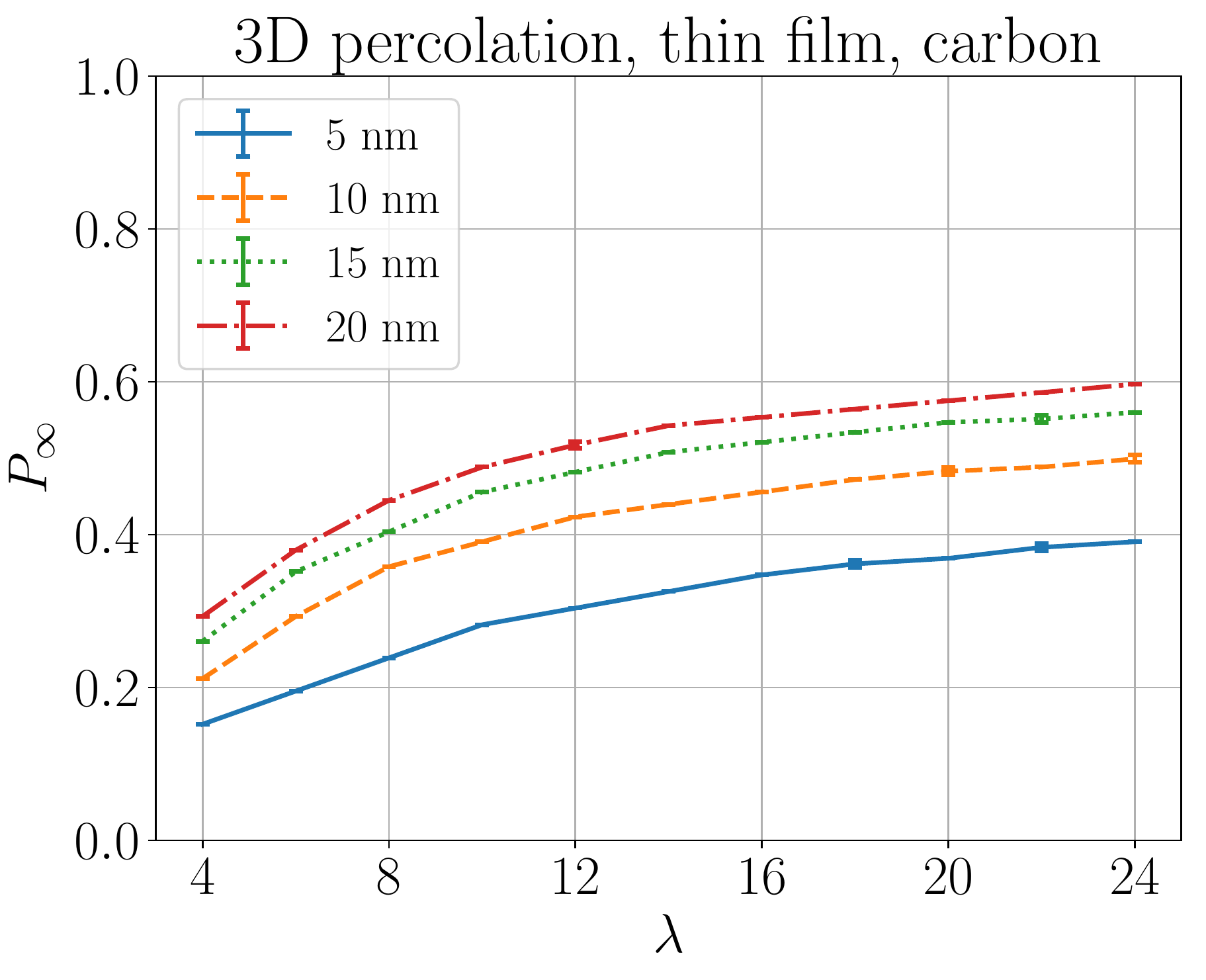}
~
\includegraphics[width=0.63\columnwidth]{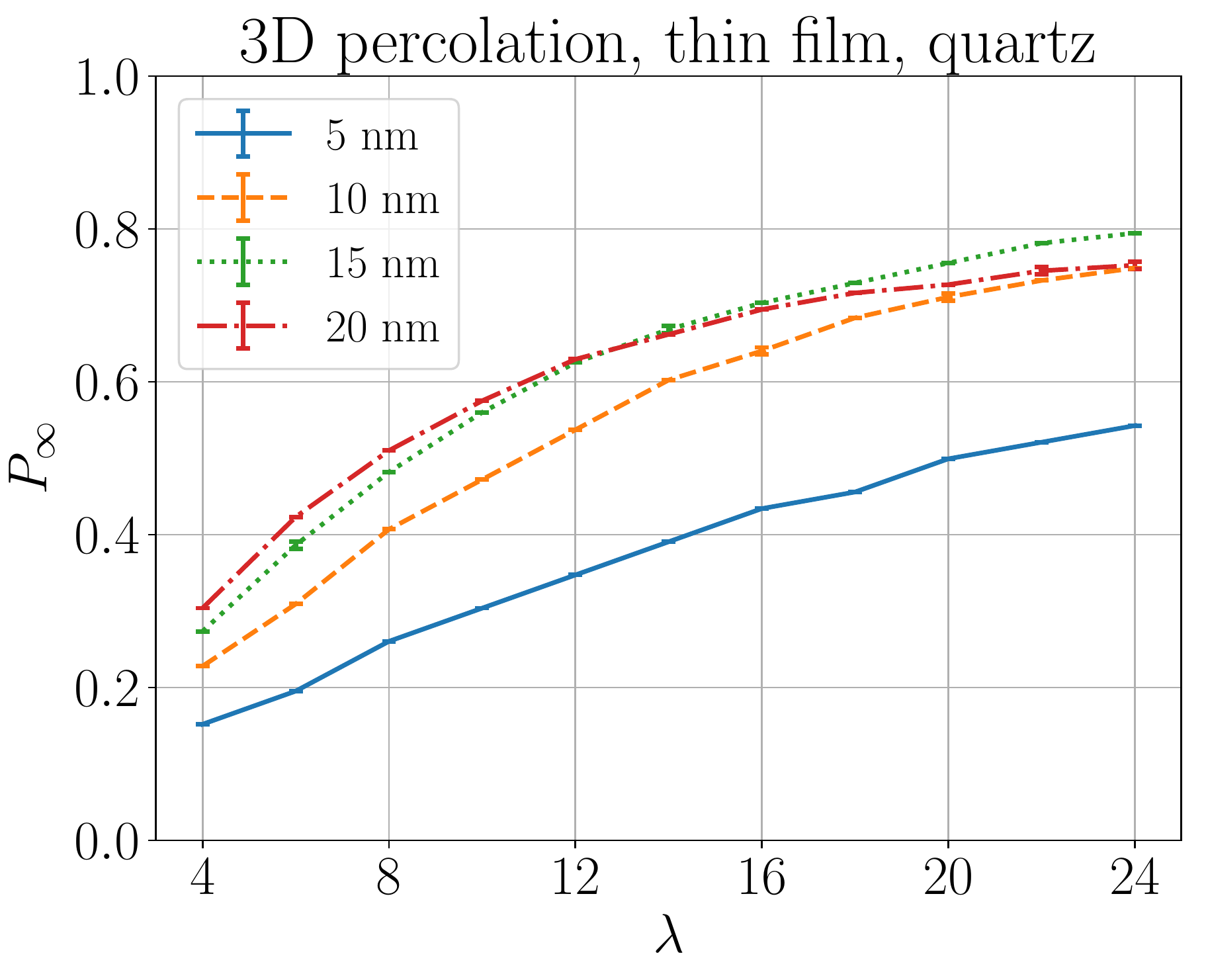}
\caption{(Color online.) Percolation cluster strength $P_\infty$ of water network in Nafion bulk and thin film confined by carbon and quartz, serving as a quantitative measure for water cluster connectivity.}
\label{fig:pinf}
\end{figure*}

The 2D and 3D water percolation cluster strengths are shown in Fig.~\ref{fig:pinf}. In case of carbon confinement, the 2D percolation does not depend strongly on the film thickness, but 3D percolation is qualitatively similar to the diffusivity curves in Fig.~\ref{fig:diff}, showing the same spacing of the equal film thickness curves. This suggests that the diffusivity of DPD beads computed in the previous section might be, after all, a good approximation for protonic conductivity. On the other hand, the 2D percolation of quartz demonstrates strong dependence on film thickness and, contrary to the intuition, the 5~nm film shows very high 2D percolation at low water uptakes. This can be confirmed by inspecting the clusters for e.g. $\lambda=4$ in Fig. 7 in the supplementary material, which shows visibly better connectivity at 5~nm than 10 or 15~nm.

Following the insights by Hsu~\cite{Hsu_MA_1980} and Kirkpatrick~\cite{Kirkpatrick_RMP_1973}, our percolation analysis shows that in order to optimise for protonic conductivity or transport properties in general, hydrophobicity of the confining substrate is a key parameter. This conclusion was also reached by recent experiments.~\cite{Orfanidi_JES_2017}

\section{Conclusion}
We have simulated Nafion thin films confined by two materials, carbon and quartz, using dissipative particle dynamics. This well-established mesoscale method enabled the use of a large box size and rapid equilibration, compared with classical molecular dynamics. We used film thicknesses likely to be found in the catalyst layer of fuel cells, ranging between 5 and 20~nm.

Our simulations show that the clustering of water and the PTFE backbone in the direction normal to the thin film is driven by the confinement scale, water content and the hydrophobicity of the confining material. The number of clusters increases with film thickness, and the cluster size depends on the water content but not the ionomer film thickness. For hydrophobic carbon, a depletion zone with little water is formed at the ionomer-carbon interface, whereas for hydrophilic quartz, water accumulates at the quartz-ionomer interface. These findings are in accord with the experiments performed by the NIST group.~\cite{DeCaluwe_SM_2014,Dura_MA_2009,Kim_MA_2013}

Percolation analysis of water in the thin ionomer films reveals patterns in cluster size and connectivity that change with the confining material. Both carbon and quartz establish a well-connected network of channels. Water diffusivity shows significant anisotropy, regardless the of confining material. The liquid moves up to 20\% more readily in the direction parallel to the thin film, compared to in the normal direction. This anisotropy increases with decreasing film thickness.

Our findings offer a perspective on the role of surface hydrophobicity of electrode materials deployed in the catalyst layer of fuel cells, and the direction of the flow of water formed on catalyst nanoparticles from protons and oxygen.

\section{Supplementary material}
Supplementary material contains the VMD screenshots, water and PTFE profiles, and clustering of all the explored configurations.

\section{Acknowledgments}
The authors thank Rob Groot and Dash Fongalland for useful discussions and Boris Fa\v{c}kovec for proofreading the manuscript. P.V. and J.A.E. acknowledge the support of EPSRC and Johnson Matthey.

\bibliography{ref.bib}

\begin{thebibliography}{49}
\bibitem [{not()}]{note_wc}%
  \BibitemOpen
  \href@noop {} {\enquote {\bibinfo {title} {Qualitatively, it does not matter
  if it is a water content or volume fraction of water.}}\ }\BibitemShut
  {NoStop}%
\end{thebibliography}

\end{document}